\documentclass[11pt,a4paper]{article}
\pdfoutput=1

\usepackage{jheppub}

\usepackage{epsfig,multicol,bbm}
\usepackage{amsmath,amsfonts,amssymb,longtable,mathtools}
\usepackage{array}
\usepackage{graphicx}
\usepackage{subfig}
\usepackage{enumerate}

\def\bea{\begin{eqnarray}}
\def\eea{\end{eqnarray}}

\input epsf.sty

\usepackage[makeroom]{cancel}
  \usepackage{latexsym}
  \usepackage{epsf}
  \usepackage{amssymb}
  \usepackage{graphicx}
  \usepackage{amsmath}
  \usepackage{amsmath,amssymb,amsthm}
  \usepackage{verbatim}

\renewcommand{\d}{\textrm{d}}

\def\ni{\noindent}
\def\bi{\begin{itemize}}
\def\ei{\end{itemize}}
\def\be{\begin{equation}}   
\def\ee{\end{equation}}
\def\ben{\begin{equation*}}
\def\een{\end{equation*}}
\def\c1{C_1}
\def\d1{D_1}

\begin{document}
\preprint{MI-TH-1763}
\title{D-brane Disformal Coupling and Thermal Dark Matter}

\author{Bhaskar Dutta\footnote{dutta@physics.tamu.edu}$^{a}$,}
\affiliation{${}^a$Mitchell Institute for Fundamental Physics and Astronomy, Department of Physics and Astronomy, Texas A\&M University, College Station, TX 77843, USA}
\author{Esteban Jimenez\footnote{este1985@physics.tamu.edu}$^{a}$,}
\author{Ivonne Zavala\footnote{e.i.zavalacarrasco@swansea.ac.uk}$^{b}$}
\affiliation{${}^b$Department of Physics, Swansea University, Singleton Park, Swansea, SA2 8PP, UK}

\abstract{Conformal and Disformal  couplings between a scalar field and matter occur naturally in general scalar-tensor theories. In  D-brane models of cosmology and particle physics,  these couplings originate   from the D-brane action describing the dynamics of its transverse (the scalar) and longitudinal (matter) fluctuations, which are thus coupled. During the post-inflationary regime and before the onset of big-bang nucleosynthesis (BBN), these  couplings can modify the expansion rate felt by matter, changing the predictions for the thermal relic  abundance of dark matter particles and  thus the annihilation rate required to satisfy the dark matter content today. 
We study the D-brane-like conformal and  disformal couplings effect on the expansion rate of the universe prior to  BBN and its impact on the dark matter relic abundance and annihilation rate. 
For a purely disformal coupling, the expansion rate is always enhanced with respect to the standard one. This gives rise to larger cross-sections when compared to the standard thermal prediction for  a range of dark matter  masses, which will be probed by future experiments.  
In a D-brane-like scenario, the scale at which the expansion rate enhancement occurs depends on the  string coupling and the string scale.}


\maketitle

\section{Introduction}

Recent cosmological data support the phenomenological $\Lambda$-Cold Dark Matter ($\Lambda CDM$) model  for cosmology, that describes the energy density content  of the universe in terms of a cosmological constant, $\Lambda$ and cold dark matter -- which together make up  95\% of the universe's energy density budget -- as well as the $\sim 5\%$ made of baryonic standard model (SM) particles. 
This phenomenological model is complemented with the inflationary mechanism, the most successful  framework to date  to account for the origin of  structure in the universe. After inflation, the universe must reheat providing the initial conditions for the hot big bang. 

However, the  physics describing the universe's evolution from the end of inflation to the onset of big bang nucleosynthesis (BBN) at around $t\sim 200-300$ s ($T\sim 1$ MeV) remains  highly unconstrained, while the  predicted abundances of the light elements resulting from primordial nucleosynthesis are in very good agreement with available observational data, which strongly supports our understanding of the universe's evolution back to the first seconds after the big bang.
During this unconstrained period, the universe may have gone through a ``nonstandard"  expansion and still be compatible with BBN. If such modification happened during DM decoupling,  DM freeze-out may be modified with measurable consequences for the relic DM abundances.

In an attractive scenario of the cosmic history, the universe is radiation dominated prior to BBN and  dark matter is produced from the thermal bath created at the end of inflation. 
In this thermal picture,  the observed relic density is satisfied for  DM species with weak-scale interaction rate $\langle\sigma v \rangle$, which is around $3.0\times10^{-26}{\rm cm^3s}^{-1}$ (or $\sigma \sim 1$ picobarn), corresponding to weak interactions.
 Despite such a small value, the Fermi-LAT and Planck experiments have been exploring upper bounds on $\langle\sigma v \rangle$ (see  Refs.~\cite{Fermi-LAT:2016uux,Ade:2015xua}). In the future, HAWC~\cite{Abeysekara:2014ffg} and CTA~\cite{Carr:2015hta} will probe the annihilation rate for a wide range of dark matter masses.  
  It is thus worth establishing whether a larger or smaller annihilation rate than the standard thermal prediction, could still have a thermal origin due to modifications to the standard  cosmological evolution before BBN.

On the other hand, string theory approaches to   SM  particle physics and inflation model building generically predict the presence of several new ingredients, and in particular new particles such as scalar fields with clear geometrical interpretations. 
Type II string theory  models of particle physics introduce new ingredients such as D-branes, where matter (and DM) is to be localised\footnote{For a review on D-brane models of particle physics  see e.g. Ref.~\cite{IU}.}. In D-brane constructions, longitudinal string fluctuations are identified with the matter fields such as the SM and/or DM particles, while transverse fluctuations correspond to scalar fields, which may play a role during the cosmological evolution\footnote{For example, a coupled dark energy - dark matter D-brane scenario was proposed in Ref.~\cite{KWZ}.}. 
These scalars couple conformally and disformally to the matter living on the brane \cite{KWZ} and thus may change the cosmological expansion rate felt by matter and the standard predictions for the DM relic abundance \cite{DJZ} as well.

The  modifications to the  relic abundances   in  conformally coupled scalar-tensor theories (ST) such as generalisations of the Brans-Dicke  theory were first discussed by Catena et al.~in Ref.~\cite{Catena}. These authors showed a general enhancement on the modified expansion rate, $\tilde H$ with respect to the standard general relativity (GR) rate, $H_{GR}$, before rapidly dropping back to the GR value well before the onset of BBN. They also found that this rapid relaxation of the scalar-tensor expansion rate towards $H_{GR}$ led to a reannihilation effect:  after the initial particle decoupling, the dark matter species experienced a subsidiary period of annihilation as the expansion rate of the universe dropped below the interaction rate.
In Ref.~\cite{DJZ}, we established the conditions under which this effect happens. As we discussed there, a reannihilation phase can occur for a nontrivial set of initial conditions for suitable conformal couplings. We found that  for dark matter particles with large masses ($m\sim 10^3$GeV) the particles  undergo this second annihilation process. Moreover,  we also determined that the annihilation rate had to be up to four times larger than that of  standard cosmology  in order to satisfy the dark matter content of the universe of 27 \%. On the other hand, for smaller masses this reannihilation process does not occur, but we found that for masses of around 130 GeV, the annihilation rate can be smaller than the annihilation rate in the standard cosmological model. Further  studies  on  conformally  coupled  ST  models  have  been  performed in recent  years in Refs.~\cite{Catena2,Gelmini,RG,WI,MW} (see also Refs.~\cite{LMN,Pallis,Salati,AM,IC,MW2,MW3}).

 In scalar-tensor theories the most general physically consistent relation between two metrics in the presence of a  scalar field, is given by\footnote{More generally, the functions, $C$ and $D$ can depend on $X=\frac12 (\partial \phi)^2$ as well. We do  not consider this case in the present paper. } \cite{Bekenstein}:
\be\label{Dmetric}
\tilde g_{\mu\nu} = C(\phi) g_{\mu\nu} + D(\phi) \partial_\mu \phi \partial_\nu \phi\,.
\ee
The first term in Eq.~\eqref{Dmetric}  is the conformal transformation which
characterises the Brans-Dicke class of scalar-tensor theories widely explored in the literature   
\cite{Catena,Catena2,Gelmini,RG,WI,MW}. The second term is the so-called disformal coupling, which is generic in extensions of general relativity. In particular, it arises naturally in D-brane models, as  discussed in Ref.~\cite{KWZ} in a natural model of coupled dark matter and dark energy. 
 In Ref.~\cite{DJZ}, we studied briefly  the effect of turning on the disformal term in addition to the conformal one studied in Ref.~\cite{Catena} in a phenomenological setup. In such a case, the functions $C$ and $D$ are in principle independent functions, so long as they satisfy the causality constraints:  
  $C(\phi)>0$  and $C(\phi) +2D(\phi)X>0$ ($X=\frac12 (\partial \phi)^2$) \cite{Bekenstein}.
 However we found  that in order to have a real positive modified expansion rate, $\tilde H$, the conformal and disformal factors  need to  satisfy a nontrivial relation \cite{DJZ}.  Moreover, when turning on a small disformal deformation to the conformal case, the profile of the modified expansion rate has a similar shape with a comparable enhancement with respect to the standard expansion rate and  a possible reannihilation phase. The net  effect is the possibility to have larger and smaller annihilation rates for a large range of  masses of the DM candidate for the observed DM content.

In the present work, we study in detail the effects on the expansion rate and  the DM relic abundances of the disformal coupling in \eqref{Dmetric},  which arises in the case of matter localised on D-branes.  In this case, the conformal and disformal terms are closely related and dictated by the underlying theory, such as  type IIB flux compactifications in string theory.  
The picture we have in mind is the following. After string theory inflation reheating takes place, giving rise to a thermal universe. At this stage standard model particles and dark matter should be produced. The SM would arise from stacks of D-branes at singularities or intersecting at suitable angles \cite{IU}, while DM particles could arise from  the same or a different stack of D-branes, which may be moving towards their final stable positions in the internal six-dimensional space before the onset of BBN. 
From the end of inflation to BBN, a nonstandard cosmological evolution can take place without spoiling the predictions of BBN, in particular, a change in the expansion rate felt by the matter particles due to the D-brane conformal and disformal couplings between the scalar field(s) (associated the transverse brane fluctuations) and  matter fields (associated to the longitudinal brane fluctuations). As we will see, due to the coupling the expansion rate will generically be enhanced, allowing for DM annihilation rates larger than the standard prediction\footnote{It is interesting to notice that a phenomenological model with a faster-than-usual expansion at early times,  driven by a new cosmological species, has recently been discussed in Ref.~\cite{DEramo}. }. 
Let us stress that a  realistic string theory scenario would be more complicated and may include nonuniversal couplings to baryonic and dark matter. However, it is very interesting that scalar couplings present in string theory can give interesting modifications of the post-inflationary evolution after string inflation.

In this paper we show for the first time that   this enhancement  happens due to a purely disformal contribution or  a combination of conformal and disformal terms. The  former case -- a disformal enhancement -- is particularly interesting as it can be interpreted in terms of an ``unwarped" compactification, which is typical of a large-volume compactification of  string theory,  which is needed for perturbative control. 
When we identify the scale arising from the disformal coupling with the tension of a moving D3-brane, where matter is localised, this scale is determined by the string scale and the string coupling.  Interestingly, the modification of the expansion rate can take place at different temperature scales, depending on the value  of the string scale.  
When we turn on the conformal coupling, whose geometrical interpretation is a nontrivial warping, the profile of the modified expansion rate resembles  the disformal example studied in Ref.~\cite{DJZ}, 
allowing for a reanihilation phase for some DM masses and suitable initial conditions.
Compared to the pure conformal case \cite{Catena},  (conformally and) disformally coupled scalar-tensor theories offer a richer phenomenology.%

The enhancement of expansion due to conformal and disformal terms impact the early universe cosmology and we study dark matter phenomenology in this paper.  We use the thermal freeze-out picture as an example and study its impact on the correlation of annihilation cross section with the dark matter content. This enhancement can also affect  other cosmological phenomena, e.g., leptogenesis, which would require a detailed understanding of the  model arising from  string theory. 

The rest of the paper is organised as follows. In the next section, we first briefly introduce  briefly the general D-brane-like setup following  the conventions and notation of Ref.~\cite{DJZ} (see also Ref.~ \cite{KWZ}). We then go directly to the cosmological equations and discuss how the expansion rate is modified in general. In section \ref{Sec:3} we  move on to the D-brane-like case, where the conformal and disformal functions are related. We start discussing the equations in the Jordan frame as well as the initial conditions and constrains that we use to  numerically solve the full equations. We then discuss in detail the solutions for the unwarped case, that is,  a purely disformal effect (or $C=$const.). We  numerically compute the modified expansion rate, the enhancement factor and the effects on the relic abundance and DM annihilation rate. Next we discuss the warped case, using for concreteness the same conformal function used in Refs.~\cite{Catena,DJZ}. We also comment on the result of using other functions. 
Finally in \ref{RA} we discuss the effect on the DM relic abundances and annihilation rates. 
We conclude  in Section \ref{conclusions} with a discussion and summary of our results.

\section{D-brane Disformal Coupling}\label{Sec:2}

We start this section by outlining our setup, which (as described in the introduction)  can arise from a post-string inflationary scenario. At this stage, the universe is already four-dimensional and moduli associated to the compactification have been properly stabilised\footnote{Though these fields might be displaced from their minima, giving rise to a matter dominated regime, with interesting consequences (see e.g. Ref.~\cite{Allahverdi:2013noa}).}. However, the relevant parameters in the model will depend on the string theory quantities such as the string scale, string coupling and compactification volume as we will argue. 

The starting action we consider is given by 
\bea\label{S1A}
S=S_{EH} + S_{brane}\,,
\eea
where: 
\bea
&& \hskip -0.9cm S_{EH}=\frac{1}{2\kappa^2} \!\!\int{\!d^4x\sqrt{-g}\,R}, \\
&& \hskip -0.9cm S_{brane} = - \!\!\int{\!d^4x\sqrt{- g} \left[ M^4 C ^2(\phi)\sqrt{1+\frac{D(\phi)}{C(\phi)} (\partial\phi)^2}+V(\phi)\right]} 
 - \!\!\int{\!d^4x\sqrt{-\tilde g} \,{\cal L}_{M}(\tilde g_{\mu\nu}) } \,, \label{braneS}
\eea
where in a string setup, $\kappa^2=M_{P}^{-2}=8\pi G$ is related to the string coupling, scale and  overall compactification volume by 
$M_P^2 = \frac{2{\cal V}_6}{2\pi g_s^2 \alpha'} $, where $M_s^{-2}=\ell_s^2=\alpha'(2\pi)^2$  is the string scale,  ${\cal V}_6$ is the dimensionless six-dimensional (6D) volume in string units and $g_s$ is the string coupling. Note also   that $G$ is not in general equal to Newton's constant as measured by e.g.~local experiments. 

In   \eqref{braneS}  we describe the brane dynamics (of transverse and longitudinal fluctuations associated to the scalar and matter respectively) given by the Dirac-Born-Infeld (DBI) and Chern-Simons  actions for a single D3-brane. The DBI part gives rise to the noncanonically normalised  scalar field $\phi$, associated to the single overall position, $r^2=\sum_i^6 y_i^2$, of the brane in the  internal 6D space\footnote{In general, a D3-brane can move in all  six of the internal dimensions.}  with coordinates $y_i$.  In this case, the scale $M$ is dictated by the tension of a D3-brane as $M^4=T_3 =(g_s\alpha'^2(2\pi)^3)^{-1} = M_s^4(2\pi)g_s^{-1}=\frac{g_s^3}{8\pi {\cal V}_6^2}M_P^4$ and thus by the string scale and coupling.  In reality, one would most likely have a stack of branes moving in the internal space. However, to study the cosmological evolution after inflation, it is enough to model all matter living on the moving brane as in Eq.~\eqref{braneS} (see also Refs.~\cite{KWZ,DWZ}) via the disformally coupled matter Lagrangian ${\cal L}_M$. 

In Ref.~\eqref{braneS},  the disformally coupled metric $\tilde g_{\mu\nu}$  is given by the induced metric on the brane, which for a brane moving along a single internal direction can be written as 
\be\label{gammaA}
\tilde g_{\mu\nu} = C(\phi) g_{\mu\nu} + D(\phi) \partial_\mu \phi \partial_\nu \phi\,.
\ee
where the scalar field is related to the D-brane position by\footnote{For the D3-brane case,  one can also consider different dimensionalities, which will add extra factors due to the internal volumes wrapped by the brane in that case. } $\phi = \sqrt{T_3} \,r$, and while $C(\phi)$ is dimensionless, $D(\phi)$ has units of mass$^{-4}$. These functions are specified by the ten-dimensional (10D) compactification and therefore in general will be related to each other as we  see below (see also \cite{KWZ}). 

\subsection{The equations of motion}

Einstein's equations  obtained from  Eq.~eq\ref{S1A} are given by
\be
R_{\mu\nu} -\frac{1}{2}g_{\mu\nu} R = \kappa^2\left(T^\phi_{\mu\nu} + T_{\mu\nu}\right)\,,
\ee
where the energy-momentum tensors are defined with respect to the Einstein-frame metric $g_{\mu\nu}$ and are given by
\be\label{Tm}
T_{\mu\nu}= P g_{\mu\nu} +(\rho + P) u_\mu u_\nu\,,
\ee
for matter, where $\rho$, $P$ are the energy density and pressure for matter with equation of state $P/\rho = \omega$. For the scalar field, the energy-momentum tensor takes the form:
\be\label{EMphiA}
T_{\mu\nu}^{\phi} = - g_{\mu\nu} \left[M^4C^2 \gamma^{-1} + V \right] 
+ M^4 CD \,\gamma \partial_\mu\phi \, \partial_\nu \phi
\ee
where the energy density and pressure for the scalar field are identified as:
\bea\label{rhoPA}
\rho_\phi =M^4C^2 \gamma + V  \,, \qquad P_\phi =  - M^4C^2\gamma^{-1} - V  \,,
\eea
 and the  ``Lorentz factor"  $\gamma$ introduced above is defined by
\be\label{LorentzA}
\gamma \equiv \left(1+ \frac{D}{C}\, (\partial\phi)^2\right)^{-1/2}\,.
\ee
It will be  convenient to rewrite Eq.~\eqref{rhoPA} by introducing  ${\mathcal V} \equiv V + C^2 M^4$, as 
\be\label{rhoPA2}
\rho_\phi =- \frac{M^4C D\gamma^2}{\gamma+1} (\partial\phi)^2 + {\mathcal V}  \,, \qquad 
P_\phi = - \frac{M^4C D\,\gamma}{\gamma+1}(\partial\phi)^2 - {\mathcal V}  \,.
\ee
The equation of motion for the scalar field is:
\bea\label{EqphiA}
&&\hskip-1cm - \nabla_\mu\!\left[M^4 DC\gamma\, \,\partial^\mu \phi\right ] \!+ \!\frac{\gamma^{-1}M^4 C^2}{2} \!\left[\!\frac{D_{,\phi}}{D} +3\frac{C_{,\phi}}{C} \right] \!+ \!\frac{\gamma \,M^4 C^2}{2} \!\left[\!\frac{C_{,\phi}}{C} -\frac{D_{,\phi}}{D} \right] \!+\! V_\phi
\nonumber \\&& \hskip3cm 
- \frac{T^{\mu\nu}}{2}\!\left[\frac{C_{,\phi}}{C} g_{\mu\nu}  +\frac{D_{,\phi}}{C}\partial_\mu\phi\partial_\nu\phi\right]
+\nabla_\mu \left[\frac{D}{C}T^{\mu\nu} \partial_\nu\phi \right] =0\,, 
\eea
where $C_{,\phi}$ denotes derivative of $C$ with respect to $\phi$, and similarly for $D, V$.
Finally, the energy-momentum conservation  equation, $\nabla_\mu T^{\mu\nu}_{tot} = \nabla_\mu \left(T^{\mu\nu}_{\phi}+  T^{\mu\nu}\right)=0$, combined with the equation of motion for the scalar field allows us to define $Q$ as
\be
Q\equiv \nabla_\mu \left[\frac{D}{C} \,T^{\mu\lambda} \,\partial_\lambda \phi\right] - \frac{T^{\mu\nu} }{2} \left[\frac{C_{,\phi}}{C} g_{\mu\nu} +
\frac{D_{,\phi}}{C} \,\partial_\mu\phi \,\partial_\nu\phi\right]\,,
\ee
so that, 
$\nabla_\mu T^{\mu\nu}_{\phi}=- \nabla_\mu T^{\mu\nu}=  Q \partial^\nu \phi$ \cite{DJZ}.


\subsection{Cosmological equations}

Let us now look at the cosmological evolution. We start with a Friedmann-Rrobertson-Walker background  metric:
\be
ds^2= -dt^2 + a^2(t)dx_idx^i \,,
\ee
where $a(t)$ is the scale factor in the Einstein frame. With this metric, the equations of motion become
\bea
&& H^2 =\frac{\kappa^2}{3} \left[\rho_\phi +\rho\right]\,, \label{friedmann1A}\\
&& \dot H + H^2 = -\frac{\kappa^2}{6}\left[ \rho_\phi+ 3P_\phi +\rho +3 P \right]\,,\label{friedmann2A}\\
&& \ddot \phi +3H\dot\phi \,\gamma^{-2}  
+ \frac{C}{2D}\left( \frac{D_{,\phi}}{D}- \frac{C_{,\phi}}{C} + \gamma^{-2}\left[\frac{5 C_{,\phi}}{C}  - \frac{D_{,\phi}}{D}\right] 
  -4\gamma^{-3}\frac{C_{,\phi}}{C} \right)
 + \frac{1}{M^4CD\gamma^{3}}\, ({\mathcal V}_{,\phi}+Q_0) =0 \label{kgA} \,,  \nonumber \\
\eea
where, $H= \frac{\dot a}{a}$, dots are derivatives with respect to $t$, $$\gamma= (1-D \,\dot\phi^2/C)^{-1/2},$$ and 
\bea
Q_0 = \rho \left[ \frac{D}{C} \,\ddot \phi + \frac{D}{C} \,\dot \phi \left(\!3H + \frac{\dot \rho}{\rho} \right) \!+ \!\left(\!\frac{D_{,\phi}}{2C}-\frac{D}{C}\frac{C_{,\phi}}{C}\!\right) \dot\phi^2 +\frac{C_{,\phi}}{2\,C} (1-3\,\omega)
\right], \nonumber \\
\eea
where we have used the equation of state for matter $P=\omega\rho$.
The continuity equations for the scalar field and matter are given by
\bea 
&&\dot\rho_\phi + 3H(\rho_{\phi}+P_{\phi}) = -Q_0\dot\phi\,, \label{contA}\\
&&\dot\rho + 3H(\rho+P) = Q_0\,\dot\phi\,.\label{cont1A}
\eea
Using Eq.~\eqref{cont1A}, we can rewrite this  as 
\be\label{Q0A}
Q_0 = \rho\left( \frac{\dot \gamma}{\dot \phi\, \gamma} + \frac{C_{,\phi}}{2C}  (1-3\,\omega \,\gamma^2) -3H\omega\,\frac{(\gamma -1) }{\dot \phi}\right) \,.
\ee
Plugging this into  the  
(non)conservation equation for  matter \eqref{cont1A} gives
\be\label{conservaDMA}
\dot \rho + 3H (\rho + P\,\gamma^{2}) = \rho \left[\frac{\dot \gamma}{\gamma} + \frac{C_{,\phi} }{2C} \,\dot\phi\, (1-3\,\omega \gamma^2)\right]\,.
\ee
\subsection{Modified and standard expansion rates}

The modified expansion rate felt by matter $\tilde H$ (which will enter into the Boltzmann equation below) is given by the Jordan-frame expansion rate, defined in terms of Jordan (or disformal) frame quantities, defined with respect to the disformal metric $\tilde g_{\mu\nu}$. 
In this frame, the  Hubble parameter is given by: 
\be\label{tildeH}
\tilde H \equiv \frac{d \ln{\tilde a}}{d\tilde \tau} = \frac{\gamma}{C^{1/2}}\left[ H + \frac{C_{,\phi}}{2C}\dot \phi \right]\,.
\ee
and it is thus a function of the Einstein-frame rate $H$, the scalar field and its derivatives. The  proper time and the scale factors in the Jordan and Einstein  frames are related by
\be\label{tildea}
\tilde a = C^{1/2} a   \,, \qquad \quad d\tilde \tau = C^{1/2} \gamma^{-1} d\tau\,.
\ee
Furthermore, the energy densities and pressures in the two frames are related by
\be\label{rhodisform}
\tilde \rho = C^{-2} \gamma^{-1} \rho \,, \qquad \tilde P = C^{-2}\gamma P\,,
\ee
while the equation of state is given by 
\be
\tilde \omega = \omega \gamma^2\,.
\ee
One can check that in the Jordan frame, the continuity equation for  matter takes the standard form \cite{DJZ}:
\be\label{tilderho2}
\frac{d{\tilde \rho}}{d\tilde\tau}  +  3 \tilde H( \tilde \rho +\tilde P)  =0 \,.
\ee

To proceed further,  we next  swap time derivatives with derivatives with respect to the number of efolds, $N=\ln a/a_0$, so $dN=Hdt$. We also define a dimensionless scalar field $\varphi = \kappa\phi$. In this case, \eqref{tildeH} becomes: 
\be\label{tildeH2}
\tilde H = \frac{H\gamma}{C^{1/2}}\left[ 1 + \alpha(\varphi) \varphi' \right]\,,
\ee
where a prime  denotes  a derivatives with respect to $N$ and we have  defined 
\be\label{alphaeq}
\alpha(\varphi) = \frac{d\ln C^{1/2}}{d\varphi}\,.
\ee
Note also that in terms of $\varphi$ and  $N$ derivatives, the Lorentz factor is now given by
\be\label{gammaH}
\gamma^{-2} = 1-\frac{H^2}{\kappa^2}\frac{D}{C}\varphi'^2\,.
\ee

We  want to compare the Jordan-frame expansion rate with that  expected in GR, which is given by 
\be\label{HGR}
H_{GR}^2 = \frac{\kappa_{GR}^2}{3} \tilde \rho\,.
\ee
We can write this in terms of $H$, $\varphi$ and its derivatives as follows. 
We first write Eq.~\eqref{friedmann1A} as  (see Refs.~\cite{Catena,DJZ}):
\be\label{H2}
H^2 = \frac{\kappa^2}{3} \frac{(1+\lambda)}{B}\rho= \frac{\kappa^2}{3} \frac{C^2\gamma(1+\lambda)}{B} \tilde \rho\,,
\ee
where $\lambda = {\cal V}/\rho\, (=\tilde {\cal V}/\tilde \rho)$, 
\be \label{B}
B= 1-\frac{M^4CD\gamma^2}{3(\gamma+1)}\varphi'^2 \,,
\ee
and we have used Eq.~\eqref{rhodisform} in the second equality of Eq.~\eqref{H2}.
By inserting  Eq.~\eqref{H2} into Eq.~\eqref{HGR}, we can  write $H_{GR}$ entirely as a function of $H, \varphi, \varphi'$ as:
\be\label{HGR2}
H_{GR}^2 = \frac{\kappa_{GR}^2}{\kappa^2} \frac{C^{-2}B\,\gamma^{-1}H^2}{(1+\lambda)}\,.
\ee
Therefore, once we find a solution for $H$ and  $\varphi$, we can compare the expansion rates $\tilde H$ with $H_{GR}$ using Eqs.~\eqref{tildeH2} and \eqref{HGR2}. To measure  the departure from the standard expansion, we define the parameter:  
\be\label{xi}
\xi= \frac{\tilde H}{H_{GR}}\,.
\ee
Notice that $\xi$ can be larger or smaller than one, indicating an enhancement or reduction of  $\tilde H$ with respect to $H_{GR}$. This means that $\tilde H$ can grow during the cosmological evolution. However notice that this does not imply a violation of the the null energy condition (NEC). This is because  the Einstein-frame expansion rate $H$ is dictated by the energy density $\rho$ and pressure $p$, which obey the NEC and therefore $\dot H<0$ during the whole evolution, as it should (see Ref.~\cite{Creminelli}).

In the following section we describe the procedure to solve the system of coupled equations for $H$ and $\varphi$ derived from Ref.~\eqref{friedmann2A} and Ref.~\eqref{kgA}.

\subsection{Coupled equations for $\varphi$ and $H$}\label{Coupledeq}

The field  equations \eqref{friedmann2A} and \eqref{kgA} can be written as  
\bea
&& H' = - H \left[ \frac{3B}{2(1+\lambda)}   (1+ \omega) + \frac{\varphi'^2 M^4CD\gamma}{2}\right], \label{Hprime}\\ \nonumber \\
&&\!\!\varphi'' \left[1\!+\! \frac{3H^2\gamma^{-1} B}{M^4CD\kappa^2(1+\lambda)} \frac{D}{C} \right]  
+ 3\,\varphi'  \left[\gamma^{-2}-  \frac{ 3H^2\gamma^{-1} B \omega}{M^4CD\kappa^2(1+\lambda)} \frac{D}{C} \right]  \nonumber \\
 \nonumber \\
&&\hskip0.5cm +\frac{H'}{H} \varphi' \left[1+ \frac{3H^2\gamma^{-1} B}{M^4CD\kappa^2(1+\lambda)} \frac{D}{C} \right] + \frac{3B\gamma^{-3}}{M^4CD(1+\lambda)} \alpha(\varphi)(1-3 \,\omega\gamma^2) \nonumber \\ \nonumber\\
&& \hskip0.95cm 
  + \frac{3B\lambda\gamma^{-3}}{M^4CD(1+\lambda)} \frac{{\cal V}_{,\varphi}}{{\cal V}}  + \frac{3 H^2 \gamma^{-1} B}{M^4CD\kappa^2(1+\lambda)} \frac{D}{C}\left[ (\delta(\varphi) - \alpha(\varphi)) \,\varphi'^2 \right]   \nonumber \\
  \nonumber \\
  && \hskip1.5cm  +\frac{\kappa^2}{H^2}\frac{C}{D}\left[  \gamma^{-2}\left( 5 \alpha(\varphi) -\delta(\varphi)  \right)  +\delta(\varphi)-  \alpha(\varphi) \left(1+4\gamma^{-3}\right) \right]=0 \,, 
  \label{phiHeq}
\eea
where 
\be
\delta(\varphi) = \frac{d \ln D^{1/2}}{d\varphi} \label{deltaeq}\,.
\ee
We notice here that, contrary to the pure conformal case, we cannot eliminate the equation for $H$, and we end up with a single master equation for the scalar \cite{Catena}. Due to the disformal term, we need to consider the coupled  equations for $\varphi$ and $H$.

\subsubsection*{The cubic equation for $H$}

Below we solve the equations numerically, for which we need  the  initial conditions for $H_i$ and $(\varphi_i, \varphi_i')$. Therefore, we need to find an expression for $H$ in terms of all other quantities and in particular $\tilde \rho$. We can obtain this from the Friedmann equation  written in terms of $\tilde \rho$ in Eq.~\eqref{H2}.
Recalling that $\gamma$ depends nontrivially on $H$  (Eq.~\eqref{gammaH}) one obtains a cubic equation for $H^2$ given by\footnote{A similar equation was found in Ref.~\cite{DJZ} for the phenomenological disformal case. In that case, $A_2=-1$ and $A_3=0$. } : 
\be\label{cubic}
A_1 H^6 + A_2 H^4 + A_3 H^2 + A_4 =0 
\ee
where 
\bea
&&   A_1= \frac{D\varphi'^2}{C\kappa^2} \label{A1}  \,,\\
&& A_2 =  \frac{2M^4CD\varphi'^2}{3}-1 \,,\\
&& A_3 =  \frac{M^4C^2\kappa^2}{3} \left(  \frac{M^4CD\varphi'^2}{3} -2  \right) \,,\\
&& A_4 =\left(\frac{M^4\kappa^2C^2}{3}\right)^2 \frac{(1+\lambda)\,\tilde\rho}{M^4} \left( \frac{(1+\lambda)\,\tilde\rho}{M^4} + 2 \right) \label{A4} \,.
\eea

One of the solutions to Eq.~\eqref{cubic} can be written as
\be\label{Hdis}
H^2=\frac{1}{3 A_1}\left(-A_2+\left(A_2^2-3A_1A_3\right)\left(\frac{2}{\Delta}\right)^{1/3}+\left(\frac{\Delta}{2}\right)^{1/3}\right)\,,
\ee

\ni with 
\bea
\Delta&=&-27A_1^2A_4+9A_1A_2A_3-2A_2^3+\sqrt{\left(-27A_1^2A_4+9A_1A_2A_3-2A_2^3\right)^2 
-4\left(A_2^2-3A_1A_3\right)^3} \nonumber \\
& \equiv& L+\sqrt{L^2-4\ell^3}.
\eea 
The other two solutions can be  obtained by replacing 
$$ \left(\frac{2}{\Delta}\right)^{1/3} \to e^{2\pi i/3} \left(\frac{2}{\Delta}\right)^{1/3}
\qquad {\rm and } \qquad \left(\frac{2}{\Delta}\right)^{1/3} \to e^{4\pi i/3} \left(\frac{2}{\Delta}\right)^{1/3}\,.$$
We are interested in real positive solutions for $H^2$. These can be  identified by considering  a complex $\Delta$, that is, $4\ell^3>L^2$, which implies a condition on $\tilde \rho, \,\varphi'$, and $C$.  For this choice, the imaginary parts of $(\Delta/2)^{1/3}$ and  $\ell (\Delta/2)^{-1/3}$ cancel each other\footnote{In this case, we can write  $Z=\frac{\Delta}{2} =\frac{ L+i\sqrt{4\ell^3-L^2}}{2}$,  then  $Z\bar Z = \ell^3$ and $\frac{2}{\Delta} = \frac{\bar Z}{\ell^3} $ and thus the  imaginary parts in Eq.~\eqref{Hdis} cancel. }. 
We will use the real positive solutions  in our numerical implementations to find the initial condition for $H$.

\section{D-brane Disformal Solutions and the Relic Abundance }\label{Sec:3}

 As  we discussed in Section \ref{Sec:2}, when considering a probe D3-brane moving in a warped 10D space, which is a solution to the 10D equations of motion, $C$ and $D$ are related and given in terms of the warp factor of the geometry \cite{KWZ}. In particular,  in the normalisation where $\phi$ becomes canonically normalised once the DBI action is expanded,  $M^4CD=1$, (see Appendix C of Ref.~\cite{DJZ}). Other normalisations are possible; however the results will be equivalent. 
 Thus in this section   we study solutions for the D-brane conformally and disformally  coupled  matter with the choice above, which  implies $\delta(\varphi)=-\alpha(\varphi)$. We start by presenting the equations of motion for this case, followed by a discussion on the   constraints and initial conditions we use in our numerical analysis.  We first discuss in detail the numerical solutions for the $C=$ const or a pure disformal case, followed by the  $C\ne $ const case.  We then analyse the implications for the dark matter relic abundance and associated annihilation rate. For this we concentrate on the $C=$ const. case, since    $C\ne $ const. gives similar results to those studied in Ref.~\cite{DJZ}.

\subsection{Equations of motion and Jordan frame }

We are interested in the radiation- and matter-dominated eras during which the potential energy of the scalar field is subdominant. Therefore in what follows  we consider $\lambda\sim0$. Also, to solve the equations Eqs.~\eqref{Hprime} and \eqref{phiHeq}, we  need to write them in terms of Jordan-frame quantities $\tilde \omega = \omega \gamma^2$ and $\tilde \rho = C^{-2}\gamma^{-1} \rho$. After doing this, the coupled equations above become

\bea
&& H' = - H \left[ \frac{3 B}{2}   (1+\tilde \omega\gamma^{-2}) + \frac{\varphi'^2 }{2}\gamma\right], \label{Hprime1}\\ \nonumber \\
&&\!\!\varphi'' \left[1\!+\! \frac{3H^2\gamma^{-1} B}{ M^4 C^2\kappa^2} \right]  
+ 3\,\varphi' \gamma^{-2} \left[1-  \frac{ 3H^2\gamma^{-1} B}{M^4C^2\kappa^2}\tilde\omega\right] 
+\frac{H'}{H} \varphi' \left[1+\! \frac{3H^2\gamma^{-1} B}{ M^4 C^2\kappa^2} \right]\nonumber \\ \nonumber\\
 && \hskip0.5cm 
 - \frac{6H^2\gamma^{-1} B}{ M^4 C^2\kappa^2}\,\alpha(\varphi)\varphi'^2 
 +3B\gamma^{-3} \alpha(\varphi)(1-3 \,\tilde\omega)-\frac{2M^4C^2\kappa^2}{H^2}\left[ 2\gamma^{-3}-3 \gamma^{-2} +1 \right]\alpha(\varphi)=0 \,. \nonumber \\
  \label{phiHeq1}
\eea

Furthermore,  we also  convert derivatives with respect to  $N$  to derivatives with respect to  $\tilde N$, the  number of e-folds in the Jordan frame \cite{DJZ}. 
Using  Eq.~\eqref{tildea}, we see that
\be\label{N}
N\equiv\ln\left(\frac{a}{a_0}\right)=\tilde N+\ln\left[\frac{C_0}{C}\right]^{1/2}\,.
\ee
where $\tilde N \equiv \ln\left(\tilde a/\tilde a_0\right) $ and the subscript $``0"$ means that the quantity is evaluated at the present time. Since we are interested in expressing quantities as functions of temperature, we then use entropy conservation in the Jordan frame. Recalling that the entropy is given by $\tilde S=\tilde a\, \tilde s$, where $\tilde s= \frac{2\pi}{45} g_s(\tilde T) \tilde T^3$, $\tilde N$ can be expressed as
\be\label{Ntilde}
\tilde N =\ln\left[\frac{\tilde T_0}{\tilde T}\left(\frac{g_s(\tilde T_0)}{g_s(\tilde T)}\right)^{1/3}\right]\,.
\ee
Therefore,  derivatives w.r.t.~$N$ transform to  derivatives w.r.t.~$\tilde N$  (assuming well behaved functions) as: 
\bea\label{NtotildeN}
\varphi' =  \frac{1}{\left(1-\alpha(\varphi) \frac{d\varphi}{d\tilde N}\right)} \frac{d\varphi}{d\tilde N} \,,\qquad 
\varphi'' = \frac{1}{\left(1-\alpha(\varphi) \frac{d\varphi}{d\tilde N}\right)^3} \left( \frac{d^2\varphi}{d\tilde N^2} + \frac{d\alpha}{d\varphi}  \left(\frac{d\varphi}{d\tilde N}\right)^3\right)\,. \\ \nonumber 
\eea
To avoid clutter we  write down expressions with derivatives with respect to $N$, but it should be understood that all our numerical calculations are made using derivatives with respect to $\tilde N$. 

\begin{figure}[h!]
\centerline{
\includegraphics[width=.65\textwidth]{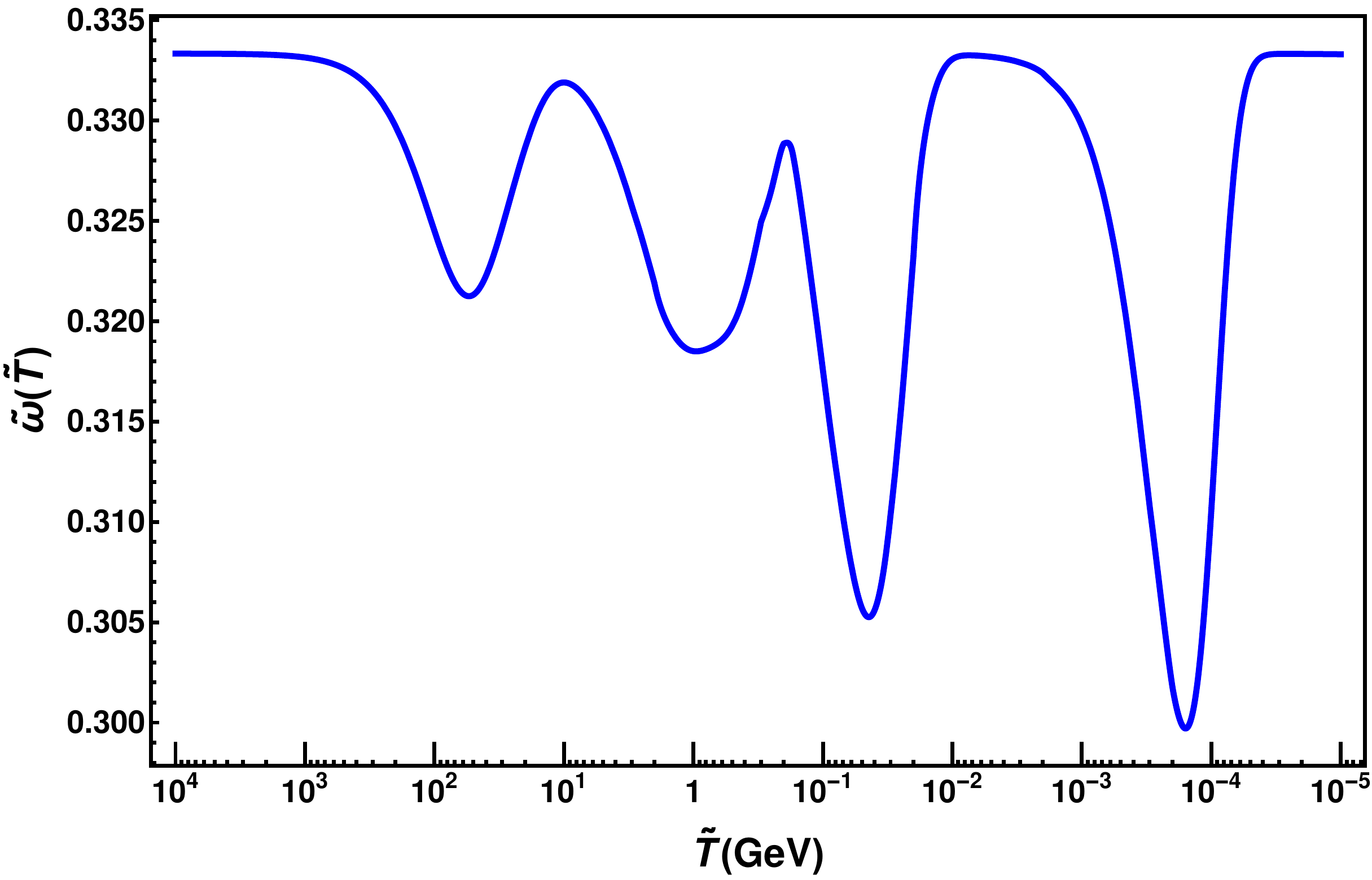}}
\caption{Equation of state $\tilde\omega$ as function of temperature during the radiation-dominated era.}
\label{plotomegaradiation}
\end{figure}

Let us start by discussing  Eq.~\eqref{phiHeq1} to understand the behaviour  of the solutions. Similarly to the  conformal case \cite{Catena,DJZ}, the derivative of $C(\varphi)$ acts as an effective potential, given by\footnote{Notice that the last term in \eqref{phiHeq1} proportional to $\alpha$ is not part of an effective potential, as it vanishes when taking the velocity terms,   $\varphi'$ to zero (so $B=1$ and $\gamma=1$).} 
\be\label{EffecV}
V_{eff}\sim  3 (1-3 \,\tilde\omega)\ln C \,.
\ee
Deep in the radiation-dominated era, the equation of state is   given by $\tilde\omega=1/3$ and the effective potential vanishes. As the temperature of the universe decreases, particle species in the cosmic soup become nonrelativistic. When the temperature of the universe drops below the rest mass of each of the particle types, nonzero contributions to $1-3\tilde \omega$  arise, activating  the effective potential. On the other hand, during the matter-dominated era,  $\tilde\omega=0$, and the effective potential is active through it.
In Section 3.1.1 of \cite{DJZ} we showed how to calculate $\tilde\omega$ during the radiation-dominated era. We reproduce here the calculation of  $\tilde \omega$ during the radiation-dominated era in Figure \ref{plotomegaradiation}.

\subsection{Initial conditions and parameter constraints }

Before we move on to solving  the coupled equations  \eqref{Hprime1} and \eqref{phiHeq1}   to find the modified expansion rate, $\tilde H$ and compare it with the standard one, $H_{GR}$, we stop here to describe  the constraints and initial conditions we use in our numerical analysis.

\subsubsection*{Parameter Constraints}\label{ParCons}

In scalar-tensor theories, deviations from GR can be parametrised in terms of the post-Newtonian parameters,  
$\gamma_{PN}$ and $\beta_{PN}$.  In the standard conformal case, these  parameters are given in terms of  $\alpha(\varphi_0)$ defined in Eq.~\eqref{alphaeq} and  its derivative, $\alpha'_0= d\alpha/ d\varphi |_{\varphi_0}$  as  \cite{Esposito,Will}
\be
\gamma_{PN} -1 = -\frac{2\alpha_0^2}{1+\alpha_0^2} \,, \qquad \beta_{PN} -1 = \frac12\frac{\alpha'_0\alpha_0^2}{(1+\alpha_0^2)^2} \,,
\ee
Solar System  tests of gravity -- including the perihelion shift of Mercury, lunar laser Ranging experiments, and the measurements of the Shapiro time delay   by the Cassini spacecraft \cite{Lunar,Shapiro,Cassini} --constrain  $\alpha_0$ to very small values of order $\alpha_0^2 \lesssim 10^{-5}$, while binary pulsar observations impose that $\alpha'_0\gtrsim -4.5$.
The strongest constraint applies to the the speed-up factor $\xi$, which has to be of order 1 before the onset of BBN \cite{Coc}.
Further, the relation between the bare gravitational constant and that measured by local experiments for  conformally coupled theories is given by \cite{Nordtvedt:1970uv}:  
\be\label{ratioG}
\kappa_{GR}^2 = \kappa^2 C(\varphi_0) [1+\alpha^2(\varphi_0)]\,.
\ee

For the phenomenological  disformal case, Solar System constraints and the ratio \eqref{ratioG} have  been studied  for constant $D$ in Ref.~\cite{ISS}. In particular, they found  $\kappa_{GR}^2 = \kappa^2 (1+3\Upsilon/2)$, where $\Upsilon \propto \varphi'^2_0$. As we will see, all solutions we found have $\varphi'=0$ at the onset of BBN. Therefore, for the constant conformal case, $\kappa_{GR}^2 = \kappa^2$. 
For the $C\ne $ const case, on the other hand, we will use the constraints on $\alpha$ above requiring that the standard expansion rate is recovered well before the onset of BBN. This is what we need to ensure that the predictions of the standard cosmological model are not modified.

\subsubsection*{Initial conditions and the  scale M}\label{ICLowM}

To find the numerical solutions, we need to fix the initial conditions for $H, \varphi, \varphi'$. Since $\varphi$ is given in Planck units, we take $\varphi_i, \varphi'_i \lesssim 1$.  To find the initial value for $H$, we need the real positive solution to Eq.~\eqref{cubic} given by Eq.~\eqref{Hdis}  for the  case $CDM^4=1$,  given in terms of the initial variables $\varphi_i, \varphi'_i , \tilde \rho_i$. 
In this case the coefficients $A_i$ simplify greatly. 

Writing as before
\be
\Delta = L+i \sqrt{4\ell^3-L^2} \,,
\ee
we now have 
\bea
&&  L = 2+2\varphi'^2 -\frac{7}{3}\varphi'^4 + \frac{2}{27} \varphi'^6 -3\varphi'^4 R  \,, \\
&& \mathbb L= 4 \ell^3 - L^2 = -\frac{\varphi'^4}{9}(1+R)\left[81\varphi'^4 R -(3+4\varphi'^2)(\varphi'^2-6)^2\right]\,,\\
&& \ell= \left(1+\frac{\varphi'^2}{3}  \right)^2 \,,\\
&& R = \frac{\tilde \rho}{M^4} \left( \frac{\tilde \rho}{M^4} +2 \right) \,.
\eea
From here it is not hard to see that $L$ can be either positive or negative and we require that  $\mathbb L>0$ for $\Delta$ to be complex, as required to find real positive solutions.  
In terms of the initial values for $\varphi'_i$\footnote{Recall that in our numerical solutions we take derivatives w.r.t. $\tilde N$, so $\varphi'_i $ should be read as  $\frac{\varphi_i'}{1-\alpha(\varphi_i)}$.}, this requirement implies 
\be\label{bound}
R\leq \frac{(3+4\varphi_i'^2)(\varphi_i'^2-6)^2}{81\varphi_i'^4}\,.
\ee
Recalling that during the radiation-dominated era the energy density is given by $\tilde\rho(\tilde T)=\frac{\pi}{30}g_{eff}(\tilde T)\tilde T^4$, once we fix $\varphi'_i$ and the initial temperature  $T_i$, the value of $M$ is fixed via Eq.~\eqref{bound}. Indeed,  Eq.~\eqref{bound}, is satisfied for $\tilde \rho_i/M^4$ in the interval 
$\left(0,- 1+\frac{2}{9\varphi_i'^2}\sqrt{(3+\varphi_i'^2)^3}\right)$. Or, in terms of $T_i$ and $\varphi_i'$, the value of $M$ lies in the interval:
\be\label{lowM}
\left[\left(\frac{3\,\pi\,g_{eff}(\tilde T_i) \,\varphi_i'^2}{-90\,\varphi_i'^2+ 20 \sqrt{(3+\varphi_i'^2)^3}}\right)^{1/4} \tilde T_i  \,, + \infty \right]\,.
\ee

\begin{figure}[h!]
\centerline{
\includegraphics[width=0.65\textwidth]{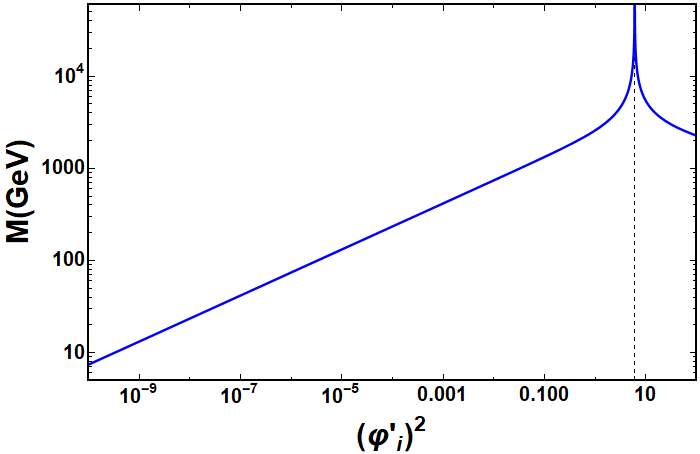}}
\caption{Lower bound for  $M$ (Eq.~\eqref{lowM}) as a function of  $(\varphi'_i)^2$ for  $C=1$. }
\label{LowerM}
\end{figure}

As an example, we show the lower bound for $M$ as a function of $(\varphi'_i)^2$ in Figure \ref{LowerM} for the initial temperature of  1.0 TeV. For simplicity we take $C=1$, so that derivatives with respect to $N$ and $\tilde N$ are the same.  As can be seen from Eq.~\eqref{lowM} and  Fig.~\ref{LowerM}, for a given initial condition $T_i$,
 the closer $\varphi'_i$ goes to $\sqrt{6}$, the larger the values of $M$, and vice versa. Also, the larger the value of $T_i$, the larger also the lower bound of $M$. In terms of the D-brane-like scenario as we described in section \ref{Sec:2}, the scale $M$ is related to the string coupling and scale  (or the six-dimensional volume) as
  $M= M_s(2\pi g_s^{-1})^{1/4}$.  Therefore, we see that the scale decreases for small string scales (large compactification volumes) and small string couplings, which are needed for the string perturbative description to be valid. We will come back to this point below.  

\subsection{Pure disformal case, $C=$ const.}\label{Ceq1}

We are now ready to discuss in detail the numerical solutions for $H$ and $\varphi$ and use them to compute the modified expansion rate. We start with  the case $C(\varphi)=$ const which can be understood as a pure disformal case, which is presented here for the first time. Indeed, notice that in this case $\gamma\ne 1$, which precisely carries the disformal (or derivative) effect, while $\alpha=0$ (which carries the conformal effect)  

Without loss of generality we can take $C(\varphi)=1$ and therefore  $D(\varphi)=\frac{1}{M^4}$. Comparing with the phenomenological case studied in Ref.~\cite{DJZ}, one could think that an arbitrary choice of the function $D$ there (with $C=1$) would give different results. However,  we expect that the effects of an arbitrary function in that case can be encoded  in the choice of the scale $M$ here, and therefore will give  similar results to those presented here.

For $C=1$, the system of coupled equations reduces to the following form,
\bea
&& H' = - H \left[ \frac{3}{2}   (1+\tilde \omega\gamma^{-2}) B+ \frac{\varphi'^2 }{2}\gamma\right]\label{HprimeC1},\\ \nonumber \\
&&\!\!\varphi'' \left[1\!+\! \frac{3H^2\gamma^{-1} B}{ M^4 \kappa^2} \right]  
+ 3\,\varphi' \gamma^{-2} \left[1-  \frac{ 3H^2\gamma^{-1} B}{M^4\kappa^2}\tilde\omega\right] + \frac{H'}{H} \varphi' \left[1+\! \frac{3H^2\gamma^{-1} B}{ M^4\kappa^2} \right]=0\,.   \label{phiHeqC1}
\eea
 As expected, the effective potential is flat, since $\alpha=0$ (see discussion above). 
We  solve these equations  numerically to find the dimensionless scalar field $\varphi$ and the Hubble parameter $H$, as  functions of $\tilde N$. 
We have explored a wide range of initial conditions for $\varphi$ and $\varphi'$ and values of the  scale  $M$. To find the initial condition for $H$ ($H_{i}$),  we use the appropriate real positive  solution of 
Eq.~\eqref{cubic}. 
We find that  at most two of the solutions Eq.~\eqref{cubic} for $H_{i}$,  are real and positive.  For these  two $H_{i}$'s, the corresponding initial value of $\gamma$ ($\gamma_i$)  is obtained using \eqref{gammaH} (setting $M^4CD=1$).  We find that one of these $\gamma_i$'s  is usually of order one while the other is 1 or 2 orders of magnitude larger (sometimes even larger). We find that  the solutions to Eqs.~\eqref{HprimeC1} and \eqref{phiHeqC1} that  obey the necessary constraints are those with $\gamma_i\sim 1$. 
Once we have found the solutions for $\varphi$ and $H$, we  go back to Eq.~\eqref{tildeH2} to obtain the expansion rate in the Jordan frame.  

Before looking into the full numerical solutions, let us take a closer look at the ratio between the modified expansion  rate and the standard rate,  \eqref{xi}. For $C=$ const this becomes 
\bea\label{xiConst}
\xi = \frac{\gamma^{3/2}}{B^{1/2}} \,.
\eea
Since $\gamma, B \geq 1$, it is clear that $\xi\geq1$, that is,  $\tilde H \geq H_{GR}$. In other words, in this case, the expansion rate is always enhanced with respect to the standard evolution. Moreover, this enhancement is driven  by $\gamma$ and $B$. As soon as $\gamma>1$, there will be a nontrivial disformal enhancement.

\begin{figure}
\centerline{
\begin{tabular}{cc}
\includegraphics[width=0.52\textwidth]{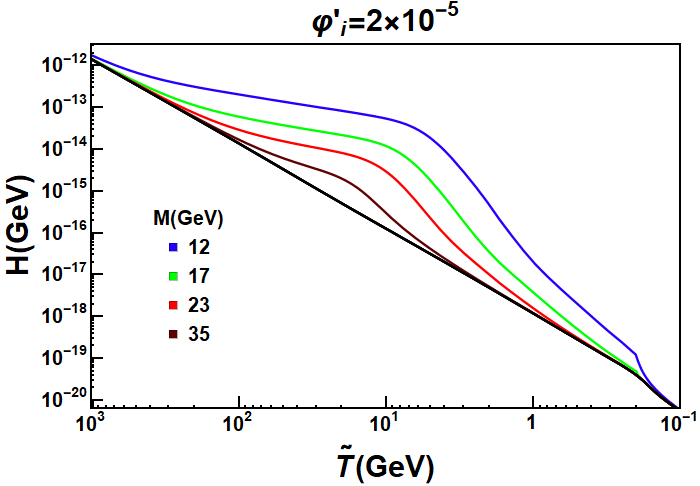}&\includegraphics[width=0.52\textwidth]{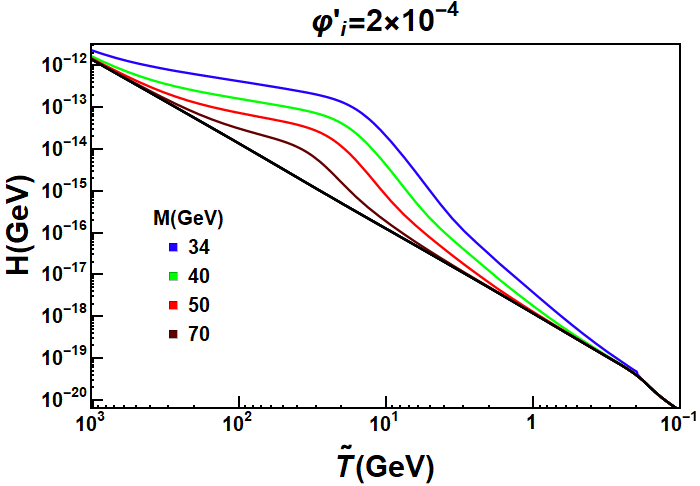}\\
\includegraphics[width=0.52\textwidth]{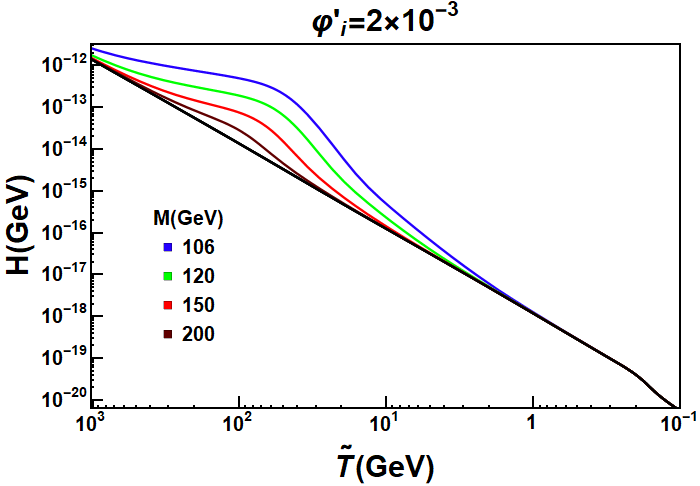}&\includegraphics[width=0.52\textwidth]{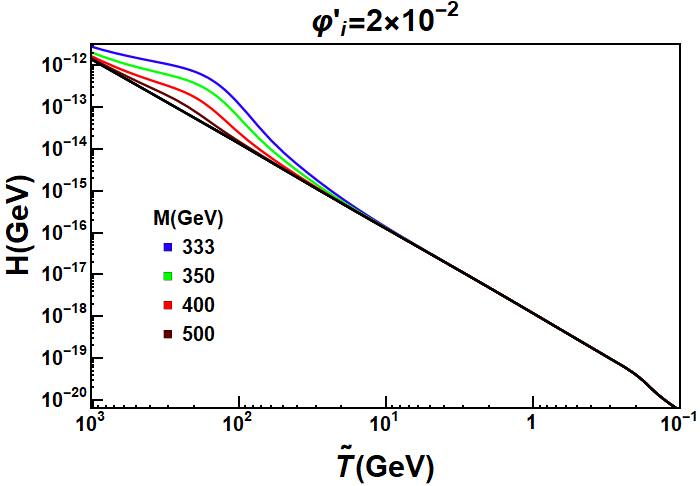}
\end{tabular}}
\caption{Modified expansion rate for the pure disformal case,  $C=1$. We show different boundary conditions and values of the scale parameter. The initial value of the scalar field for all the curves is $\varphi_i=0.2$. The black lines in all plots represent the standard expansion rate $H_{GR}$.}
\label{HubbleC1}
\end{figure}

\begin{figure}[h]
\centerline{
\includegraphics[width=0.60\textwidth]{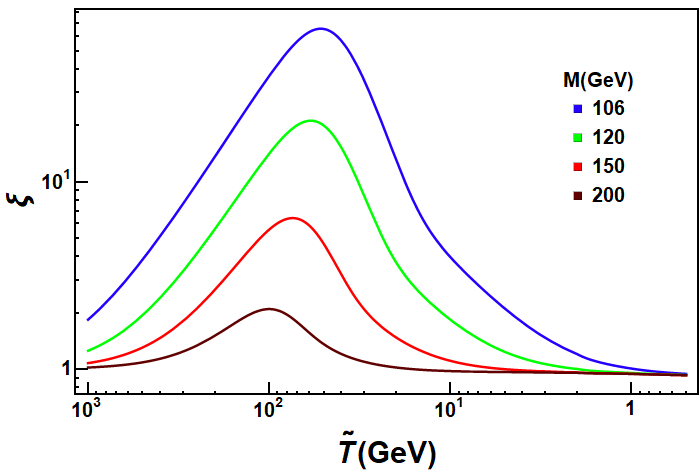}}
\caption{Speed-up factor, $\xi=\tilde H/H_{GR}$, as function of temperature for the expansion rates shown in the bottom left plot  in Figure \ref{HubbleC1}. The initial conditions chosen are $\varphi_i=0.2$ and $\varphi'_i=0.002$. }

\label{Speedup}
\end{figure}

In Figure \ref{HubbleC1} we show the resulting modified expansion rate for different values of $\varphi'$ and the mass scale $M$. In these  plots we use $\varphi_i=0.2$, but any  value in the interval $(0,1)$ and appropriate choices  of $\varphi'_i$ and the mass scale $M$, will give similar results. 
As we can see in Figure \ref{HubbleC1}, $\tilde H$ (colored lines) is always enhanced with respect to the standard expansion rate (black line),  $H_{GR}$, as discussed above. From \eqref{xiConst} and Figure \ref{Speedup}, it is clear that the ratio $\xi$  is always greater than  or equal to 1. Moreover as the temperature  decreases, the ratio $\xi$ grows from a value close to 1 (recall that  $\gamma_i\sim 1$), reaches  a maximum where $\gamma$ is maximal and  eventually decreases towards one before BBN. The  maximum value of the ratio increases and  moves to lower temperatures as the mass scale $M$ becomes smaller. 

\begin{figure}
\centerline{
\includegraphics[width=0.65\textwidth]{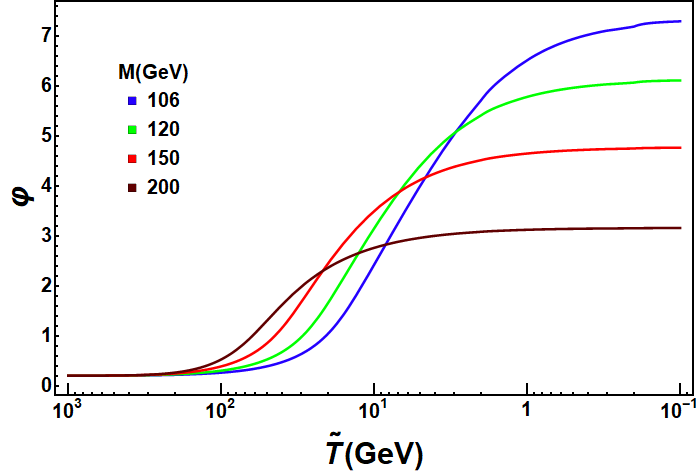}}
\caption{Scalar field as a function of temperature. The initial conditions chosen are $\varphi_i=0.2$ and $\varphi'_i=0.002$. These solutions of Eqs.~\eqref{HprimeC1} and \eqref{phiHeqC1} correspond to the expansion rates shown in the bottom left plot in Figure \ref{HubbleC1}. }
\label{phiplot}
\end{figure}

\begin{figure}[h]
\centering
\begin{tabular}{cc}
\includegraphics[width=0.5\textwidth]{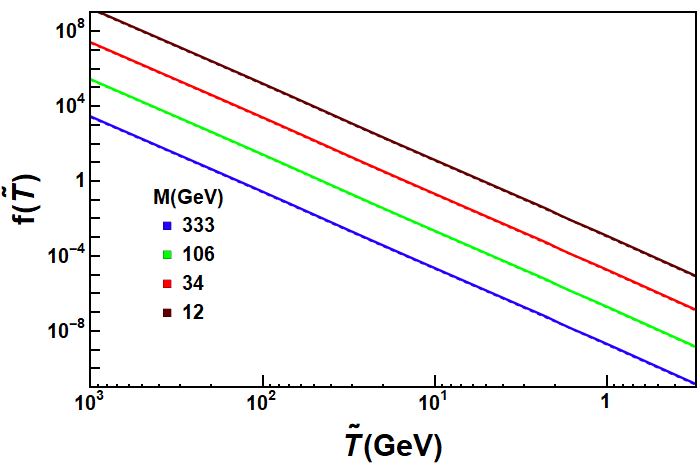}&\includegraphics[width=0.5\textwidth]{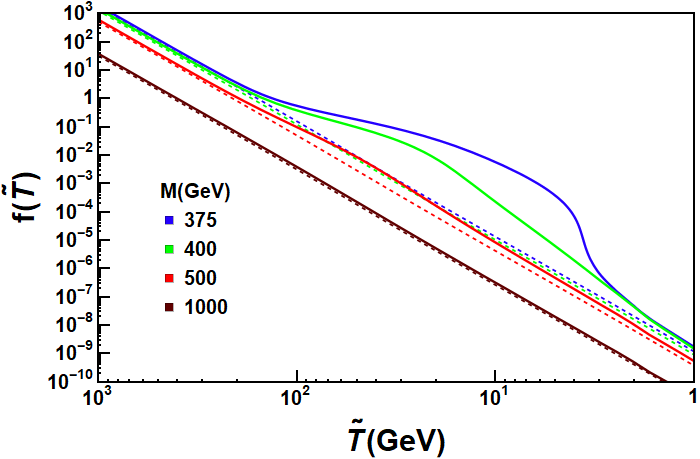}
\end{tabular}
\caption{Evolution of the factor $f$  as a function of temperature for $C=$ const case (left) and $C\ne$ const. ($f_C$, right). The initial conditions chosen in the left plot are shown in Figure \ref{HubbleC1}, while in the right plot $\varphi_i=0.2$ and $\varphi'_i=-0.004$.}
\label{factorf}
\end{figure}

We can  understand this behavior by looking at the evolution of the factor $ f=\frac{3H^2\gamma^{-1} B}{ M^4 \kappa^2}$, inside the square brackets of Eq.~\eqref{phiHeqC1}. We have seen numerically that this factor evolves as $f(\tilde T)\backsimeq \frac{3g_{eff}(\tilde T)}{10}\left(\frac{\tilde T}{M}\right)^4$ as temperature decreases (see Figure \ref{factorf}). For the  scale $M$ and temperatures plotted in Figure \ref{HubbleC1}, $f(\tilde T)$ starts  much bigger than one (up to $f(\tilde T_i)\backsimeq10^9$) and  decreases  as the temperature decreases. The bigger the  scale $M$, the earlier $f(\tilde T)$ becomes of order 1. While $f(\tilde T)$  is bigger than 1,  $\xi$ increases, the velocity of the scalar field $\varphi'$ increases slowly, and thus the scalar field increases very slowly too (see Figure \ref{Speedup} and \ref{phiplot}). As $f(\tilde T)$  approaches 1,  $\xi$ reaches a maximum and the scalar field starts  increasing faster. Then, as the temperature decreases further, $f(\tilde T)$ becomes smaller than 1. Meanwhile, $\tilde H$ starts converging  towards $H_{GR}$ (that is, $\xi$ starts decreasing) while  the scalar field keeps increasing. Finally, when $\tilde H$ becomes of order $H_{GR}$, $f(\tilde T)$ is much smaller than 1 and the scalar field  starts moving towards  a final constant value.


We see the behaviour described above in Figures \ref{Speedup} and \ref{phiplot}. 
For instance, for the  plot corresponding to  $M=106$GeV, $f(\tilde T)$ is approximately $ \frac{3g_{eff}(\tilde T)}{10}\left(\frac{\tilde T}{106 \,\text{GeV}}\right)^4$, which becomes 1 at around $\tilde T=50 \,\text{GeV}$. Between $1000$ GeV and $50$ GeV, $\tilde H$ differs from $H_{GR}$ and in this range the scalar field increases very slowly, looking almost constant. For lower temperatures, between $50$ GeV and $1$ GeV, $\tilde H$ converges towards $H_{GR}$ and the scalar field increases faster while for temperatures smaller than 1 GeV, $\tilde H\sim H_{GR}$ and the scalar field reaches its final value.

All the cases shown in Figure \ref{HubbleC1} satisfy the constraints discussed in Section \ref{ParCons}.  In particular, $\varphi'_{BBN} = 0$ (so $\Upsilon=0$)  
and the speed-up factor $\xi$ is equal to 1 prior to BBN as shown in Figure \ref{Speedup}. For  scales $M$ smaller than 10 GeV  the last condition is not satisfied, that is $\xi >1$ by the onset of BBN. Therefore,  scales  $M$ smaller than 10 GeV are discarded.

\begin{figure}
\centerline{
\begin{tabular}{cc}
\includegraphics[width=0.52\textwidth]{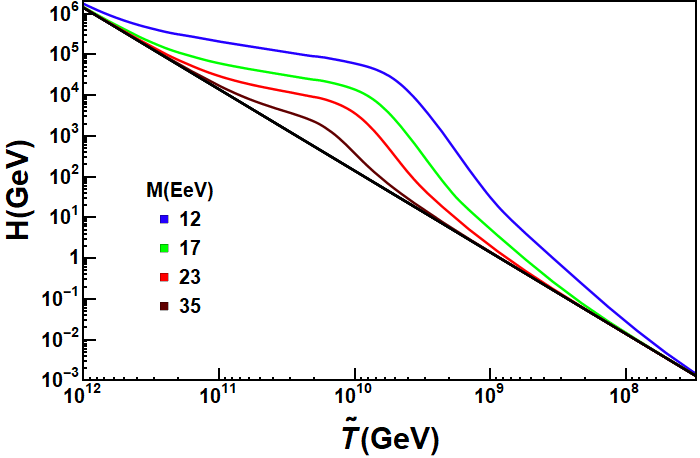}&\includegraphics[width=0.52\textwidth]{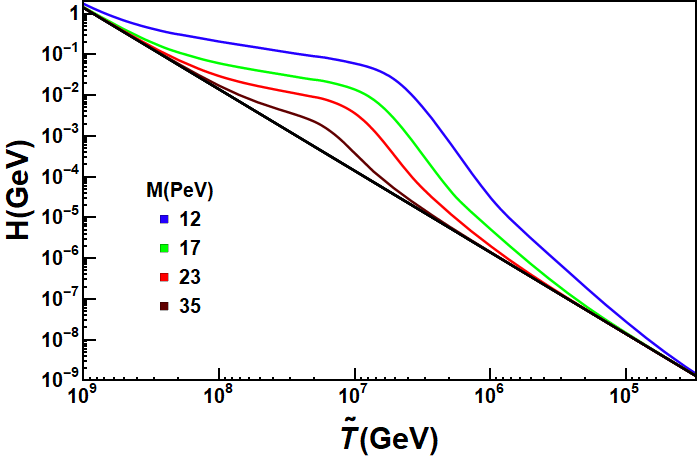}\\
\includegraphics[width=0.54\textwidth]{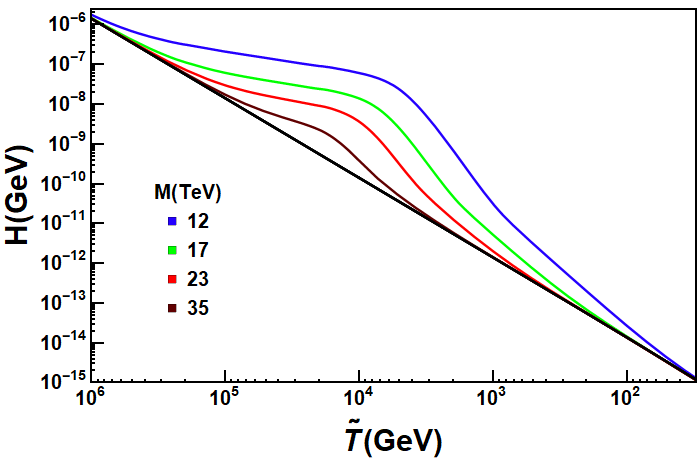}&\includegraphics[width=0.54\textwidth]{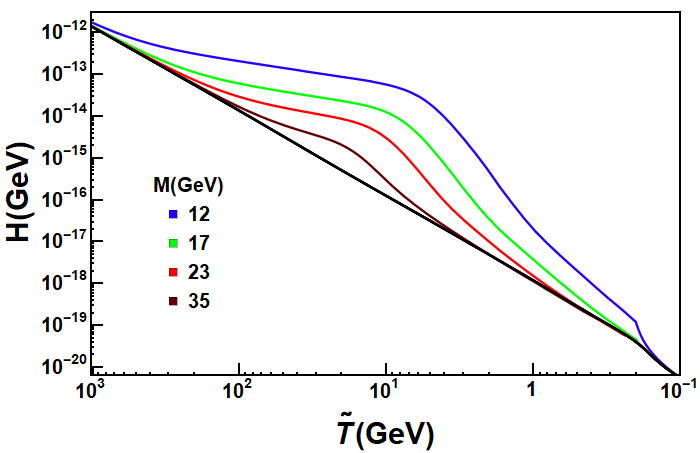}
\end{tabular}}
\caption{Modified expansion rate for the pure disformal case, $C=1$, for larger values of $M$ as compared to Fig.~\ref{HubbleC1}. For these  plots, $\varphi_i=0.2$ and $\varphi'_i=2\times10^{-5}$.}
\label{HubbleScaleM}
\end{figure}

As we have mentioned,  if we consider larger values of  $M$  than the ones presented in Figure \ref{HubbleC1}, 
the enhancement of the expansion rate will occur earlier at higher temperatures, such that $f(\tilde T)$ is  much bigger than 1 at around the initial value of the temperature, $\tilde T_i$.  To achieve this, one has to consider $M$ smaller than $\tilde T_i$, which %
happens when the initial value of $\varphi'$ is much smaller than 1. We illustrate this  in Figure \ref{HubbleScaleM} were we show a series of plots were the mass scale takes values up to order EeV. 
This figure also shows that the speed-up factor \eqref{xi},  has the same behavior as long as the ratio $\tilde T_i/M$ does not change. For instance, for all the green lines $\tilde T_i/M$=58.8. 

\subsection{Conformal and disformal case $C\neq $ const } \label{Cne1}

We now move  to the  case where the conformal coupling is not constant, so both conformal and disformal effects are turned on.  For concreteness we  consider the same conformal coupling as that studied in Refs.~\cite{DJZ} and \cite{Catena}, which is given by
\be\label{Cfactor}
C(\varphi) = (1+ b \,e^{-\beta\, \varphi})^2\,,
\ee
with  the values $b=0.1$, $\beta=8$. 
We have also analysed other functions such as $C=(b\,\varphi^2+c)^2$ with $b=4,8,15$, $c=1$. However it is harder to find numerical solutions for this and other functions, which satisfy the phenomenological constraints. In the cases  we analysed, the effect on the expansion rate  $\tilde H$ was smaller with respect to the case in Eq.~\eqref{Cfactor}. 

As  mentioned in Section \ref{Coupledeq}, the conformal term acts as an effective potential, or force, in Eq.~ \eqref{phiHeq1}, given by Eq.~\eqref{EffecV}. This effective force can be neglected when the factor $f_C=\frac{3H^2\gamma^{-1} B}{ M^4 C^2\kappa^2}$ is much larger than 1, as can be seen from Eq.~\eqref{phiHeq1}. In this regime, the evolution of the scalar field is given   by a flat effective potential,  and the scalar field stays  approximately  constant. 
When $f_C$ is becomes of order 1 or smaller and $\tilde\omega\neq 1/3 $, the evolution of the scalar field is driven  by the effective potential \eqref{EffecV} and by the Hubble friction term. 

For the conformal coupling  \eqref{Cfactor}, the effective potential allows for an interesting behaviour, according to the choice of initial conditions  \cite{DJZ}. That is, for negative initial velocities, $\varphi'_i<0$,  the scalar   field will start rolling up the effective potential towards smaller values. After reaching a maximum point, it will  turn back down the effective potential, eventually reaching its final value. 
This behaviour in the scalar field sources a nontrivial behaviour in  $C$  and (importantly) its derivative $\alpha$, and therefore in the modified expansion rate $\tilde H$. 
Indeed, when $C\ne $ const. we have 
\be\label{xiC}
\xi = \frac{\kappa}{\kappa_{GR}} \frac{C^{1/2}\gamma^{3/2} }{B^{1/2}}\left[ 1+\alpha(\varphi) \varphi'  \right]  \,.
\ee
It is not hard to see that for the initial conditions above, due to the factor inside the parentheses, $\xi$ can become less than one during the evolution.  Recalling that   $\xi=\tilde H/H_{GR}$, $\xi<1$ implies that $\tilde H<H_{GR}$,  as shown in the explicit solutions below.  This effect gives rise to the possibility of a reannihilation period, as was discussed in Ref.~\cite{DJZ}  and first pointed out in Ref.~\cite{Catena}. 

\begin{figure}
\centerline{
\includegraphics[width=0.65\textwidth]{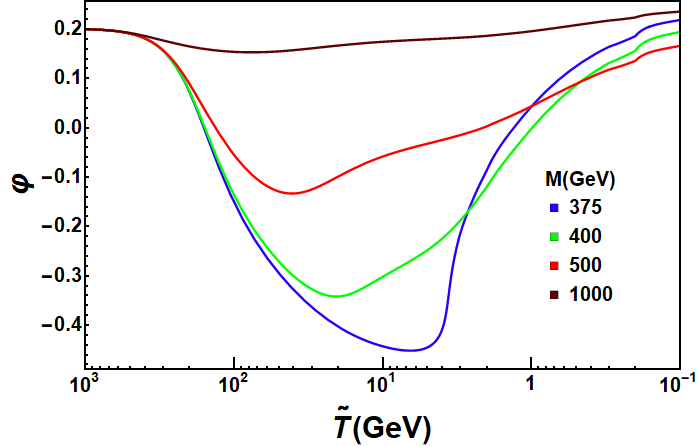}}
\caption{Scalar field as a function of temperature for different values of $M$. The conformal coupling is $(1+ 0.1 \,e^{-8\, \varphi})^2$ and the initial conditions chosen are $\varphi_i=0.2$ and $\varphi'_i=-0.004$. These solutions of Eqs.~\eqref{Hprime1} and \eqref{phiHeq1} correspond to the expansion rates shown in the right plot of Figure \ref{HubbleCnon1}. }
\label{phiplotC}
\end{figure}

Let us now take a closer look at the evolution of $f_C$ with temperature. Numerically, we  found that when $f_C \gtrsim 1$  it behaves  as  $f_C(\tilde T)\backsimeq \frac{3g_{eff}(\tilde T)}{10}\left(\frac{\tilde T}{M}\right)^4$. But when  $f_C <1$ then it evolves as  $f_C(\tilde T)\backsimeq h(\tilde T) \frac{3g_{eff}(\tilde T)}{10}\left(\frac{\tilde T}{M}\right)^4$, where $h(\tilde T)$ is a function that measures the enhancement of $\tilde H$,  which is larger than 1 and depends on the scale $M$ (see right plot in Figure \ref{factorf}). 
When $f_C\gg1$, the effective force is negligible and the scalar field  stays roughly constant. As $f_C$ decreases and approaches and/or becomes smaller than 1, the effective force takes over the evolution of the scalar field. The velocity of the scalar field starts decreasing (we use small negative velocities), and for suitable values, the scalar field goes up the effective potential and comes back down again as described above.

\begin{figure}
\centering
\begin{tabular}{cc}
\includegraphics[width=0.50\textwidth]{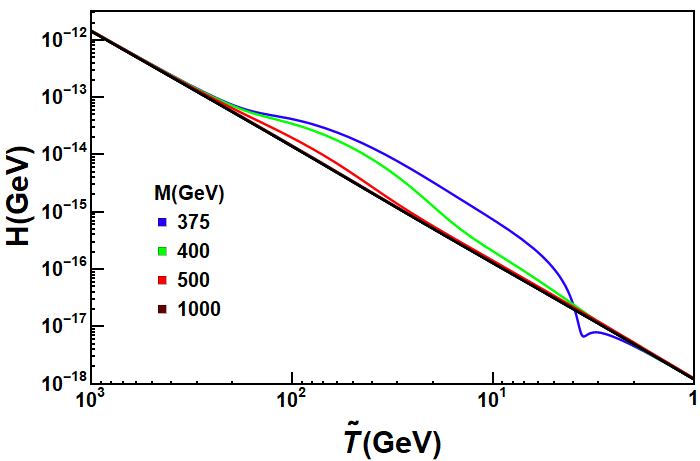}&\includegraphics[width=0.50\textwidth]{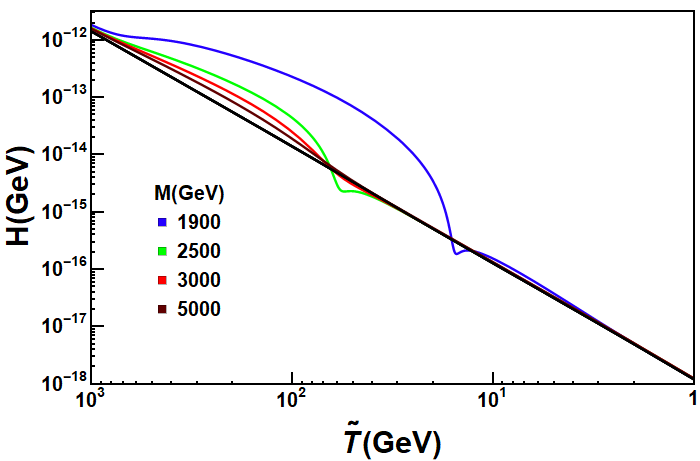}
\end{tabular}
\caption{Modified expansion rate for the case $C=1+0.1e^{-8\varphi}$.  The initial value of the scalar field for all the curves is $\varphi_i=0.2$. Also, $\varphi'_i=-0.004$ for the plot on the left and $\varphi'_i=-0.4$ for the plot on the right.}
\label{HubbleCnon1}
\end{figure}

In Figure \ref{phiplotC} we  plot the full numerical solution for  the scalar field for $\varphi_i=0.2$ and an initial velocity $\varphi'=-0.004$.  
The red, green and blue curves (scale masses smaller than  $\tilde T_i=1000$ GeV)  show the scalar field going up the effective potential toward smaller values of the field, and then rolling down  its terminal value., while for the brown curve ($M=$1000 GeV), the scalar field  stays almost constant because for this value of $M$ its initial velocity is not negative enough to move the field up the effective potential.

The effect of the scalar field on the modified expansion rate is shown in Figure \ref{HubbleCnon1} (the black straight line is $H_{GR}$). The left plot shows $\tilde H$ corresponding to the scalar field solutions in Figure \ref{phiplotC}. For these solutions, the factor $f_C$ is initially much bigger than 1 and as the temperature decreases it passes  one (around 200 GeV) and  keeps decreasing to very small values. For some values of $M$,  the scalar field goes up and  down the effective potential, producing the enhancement and the little notch in $\tilde H$ (blue), where $\xi<1$ as explained above. 
On the other hand, in the right plots, $f_C$ is initially of order 1 and then decreases to negligible values. The  initial velocity used ($\varphi'=-0.4$) is sufficiently negative to produce  the enhancement and notch in $\tilde H$ for some of the $M$ values (green and blue). 

Let us mention another point about the right plot in Figure \ref{HubbleCnon1}. For the brown curve corresponding to $M=5000\, \text{GeV}$, the enhancement is very small, and since the factor $f_C$ decreases as the mass scale $M$ increases, choosing larger values of $M$  would give a similar result, for the same choice of initial conditions. 
Indeed, as $M\to \infty$, $f_C=\frac{3H^2\gamma^{-1} B}{ M^4 C^2\kappa^2}\to 0$ and we recover the pure conformal case in Eq.~\eqref{phiHeq1}. Notice that  the last term in this equation vanishes when $M$ increases,  since $\gamma\to 1$  as $M$ increases. So, by dropping all terms proportional to $f_C$ and the last term in Eq.~\eqref{phiHeq1}  ones recovers the conformal case equations studied in Refs.~\cite{DJZ,Catena}.
For a very large value of $M$, we will recover the results of Ref.~\cite{DJZ,Catena}, by suitably changing the initial conditions for $\varphi_i$ and $\varphi'_i$.

Let us finally comment on the differences between the present case (conformal plus disformal) and the pure disformal and pure conformal cases, where there is no derivative interaction.  We saw in the previous subsection that in the pure disformal case the enhancement in the expansion rate can be produced at any  temperature (see Figure \ref{HubbleScaleM}), by suitably changing the value of the scale $M$. However, in the present $C\neq 1$ case, this does not happen at any scale since $\omega\ne1/3$ is needed and we get $\omega\ne1/3$ when SM particles become nonrelativistic.   %
This is due to the last term in Eq.~\eqref{phiHeq1}, which makes the evolution of $\varphi'$ go to zero very fast, effectively making $\gamma\sim 1$ throughout the evolution and thus an ineffective disformal enhancement. 
However, new physics at a higher scale causing a change in $\tilde \omega$ can introduce an enhancement at that scale. This will be similar to the case with the  additional $M$ scale associated with the D-brane models
The conformal  enhancement is effective so long as the effective potential \eqref{EffecV} is active, that is, whenever $\omega\ne1/3$ (see Figure \ref{plotomegaradiation}). 

\subsection{Effect on the relic abundances and cross section}\label{RA}

Now that we have computed the modified expansion rate, we move on to discuss its  impact and implications on the dark matter relic abundance and cross section. We focus on the case $C=$ const. since  the $C\ne $ const. case gives similar results to those studied in \cite{DJZ}, as we discuss below. 

For a dark matter species $\chi$ with mass $m_\chi$ and a thermally averaged annihilation cross section $\langle \sigma  v \rangle$ (where $ v$ is the relative velocity), the dark matter number density $ n_\chi$ evolves according to the Boltzmann equation 
\be\label{Boltzn}
\frac{d n_\chi}{dt} = -3 \tilde H n_\chi - \langle \sigma v \rangle \left( n_\chi^2  - (n_\chi^{eq})^2 \right)\,,
\ee
where $\tilde H$ is  the expansion rate  in the Jordan frame computed in the previous section (see Figures \ref{HubbleC1} and \ref{HubbleCnon1}), felt by the matter particles and $ n_\chi^{eq}$ is the equilibrium number density.

\begin{figure}[h]
\centerline{
\includegraphics[width=0.80\textwidth]{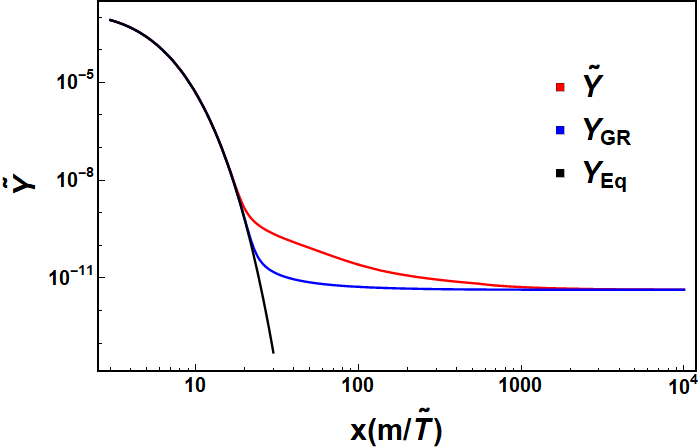}}
\caption{Abundance $\tilde Y$ for a dark matter particle  with a mass of 100 GeV. }
\label{Y100}
\end{figure}

To solve Eq.~\eqref{Boltzn}, we  rewrite it in the  standard form  in terms of $x=m_\chi/\tilde T$, 
\be\label{Boltzy}
\frac{d Y}{dx} = - \frac{\tilde s \langle \sigma v \rangle }{x \tilde H}  \left( Y^2  - Y_{eq}^2 \right) \,,
\ee
where $ Y = \frac{ n_\chi}{\tilde s}$ is the abundance and $\tilde s= \frac{2\pi}{45} g_s(\tilde T) \tilde T^3$ is the entropy density.
As a  concrete example, we solve  Eq.~\eqref{Boltzy} numerically, for the expansion rate corresponding to $M= 12$ GeV shown in the top left plot of Figure \ref{HubbleC1} and for dark matter particles with masses ranging from 10 GeV to 5000 GeV. Other choices of $M$ would give similar results. In Figure \ref{Y100} we show the solution for a DM particle of mass $m_\chi=100$ GeV. In this plot, we also include the abundance $Y_{GR}(x)$ calculated in the standard cosmology model and the abundance when dark matter particles are in thermal equilibrium, $Y_{Eq}(x)$. 

\begin{figure}[h]
\centerline{
\includegraphics[width=0.80\textwidth]{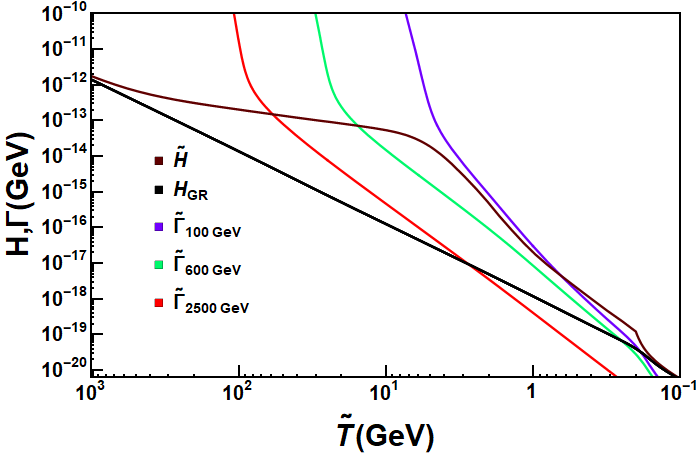}}
\caption{Expansion rate corresponding to $M=12$ GeV shown in the top left plot of Figure \ref{HubbleC1}, and interaction rates of 100 GeV (purple), 600 GeV (green) and 2500 GeV (red) GeV DM particle masses as a function of temperature. The interaction rate $\tilde \Gamma$  is given by $ \langle \sigma v \rangle\tilde s \tilde Y$.  }
\label{HGamma}
\end{figure}

In the plot we can see  how the modification of the expansion rate gives rise to an earlier than standard freeze-out (see Figure \ref {Y100} around $x=20$). This is due to the enhancement of the expansion rate $\tilde H$. As the temperature decreases ($x$ increases), $\tilde H$ becomes comparable to the interaction rate\footnote{The interaction rate is defined as $\tilde\Gamma\equiv\langle\sigma v \rangle\,\tilde s\,\tilde Y$.}  $\tilde \Gamma$ and for a small period, between $x=20$ and $x=1000$,  the abundance decreases slowly until it becomes constant.
It is interesting to notice that a similar behaviour was found in Ref.~\cite{DEramo} in a phenomenological model where an extra scalar species drives a faster than usual expansion rate, giving rise to a similar behaviour in the relic abundance. 
The comparison between $\tilde H$ (brown) and $\tilde \Gamma$ (purple) can be seen  in Figure \ref{HGamma}. 
Between around 5 GeV ($x=20$) and 0.1 GeV ($x=1000$), $\tilde H$ and $\tilde \Gamma$ are close to each other as temperature decreases. 

In  Fig.~\ref{HGamma}, we also show the interaction rate for  two other DM particle masses, 600 GeV (green) and 2500 GeV (red). Notice that for the three masses shown, once the interaction rate becomes smaller than the expansion rate $\tilde H$ (brown), it  always stays smaller than it. Therefore, there is no reannihilation effect, as we anticipated in section \ref{Ceq1}. However reannihilation can occur for the $C\ne $ const. case, where  after the first freeze-out $\tilde \Gamma$ can overcome $\tilde H$ due to $\xi <1$,  and later become smaller again. 

Let us now turn to the dark matter cross sections we have used when solving the Boltzmann equation \eqref{Boltzy}. For this we  used the observed dark matter density $\Omega_0=0.27$ to determine the thermally averaged annihilation cross section, $\langle\sigma v \rangle$ required to match the current 27\% DM content. The present dark matter content of the universe is determined by the current value of the relic abundance. This  can be obtained from the current value of the energy density parameter
$\Omega_0=\frac{\rho_0}{\rho_{c,0}}=\frac{m\,Y_0\,s_0}{\rho_{c,0}}$. Here $\rho_{c,0}$ and $s_0$ are the well-known current values of the critical energy density and the entropy density of the universe, respectively. 

\begin{figure}[h]
\centerline{
\includegraphics[width=0.65\textwidth]{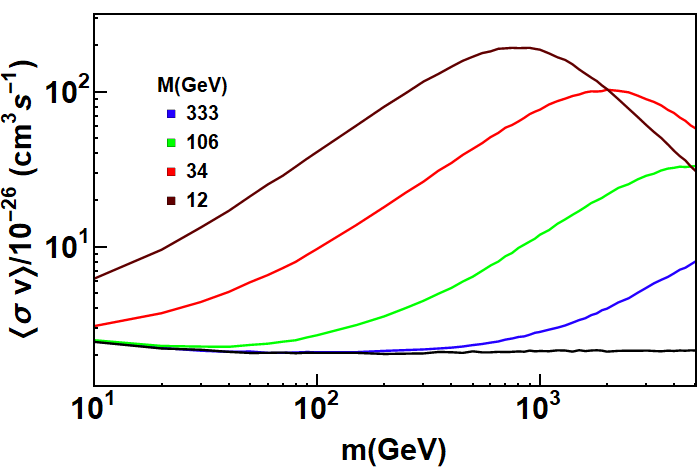}}
\caption{$ \langle  \sigma  v \rangle$ as function of dark matter particle mass. $\langle\sigma v \rangle_{GR}$ predicted by the standard cosmology model correspond to the black line, while the colored lines correspond to  $\langle\sigma v \rangle$ predicted by using the expansion rates (shown in Figure \ref{HubbleC1}), representing mass scales of $M= 12 $  (brown), 34  (red), 106  (green) and 333 GeV (blue).   }
\label{Xsections}
\end{figure}

The resulting  annihilation cross sections we determine in this way  are shown in Figure \ref{Xsections}  for dark matter masses between 10 GeV and 5000 GeV, for different values of $M$ and corresponding expansion rates $\tilde H$ shown in Figure \ref{HubbleC1}. We compare this to the annihilation cross sections  $\langle\sigma v \rangle_{GR}$ predicted by the standard cosmology model (black line), which is around $2.1 \times 10^{-26} cm^3/s$. 

The behaviour of the cross section $\langle\sigma v \rangle$ in Fig.~\ref{Xsections} shows an enhancement with respect to the standard case, with a maximum that moves towards larger dark matter masses as the scale $M$ increases.  Therefore, the smaller the  scale $M$ the larger the annihilation cross section $\langle\sigma v \rangle$. We can correlate  this behaviour with that of $\xi$ in Fig.~\ref{Speedup}, which shows the enhancement of the expansion rate. 
For example, for a mass scale of 34 GeV (red) the maximum  $\langle\sigma v \rangle$ is  around $100 \times 10^{-26} {\rm cm^3/s}$ for a DM mass of 2000 GeV,  for a mass scale of 12 GeV (brown)  the maximum  $\langle\sigma v \rangle$ is around $200 \times 10^{-26} cm^3/s$ for a DM mass of 700 GeV.
For a 600 GeV DM mass  the ratio $\langle\sigma v \rangle/\langle\sigma v \rangle_{GR}$ is almost 1 for $M=333$ GeV (blue line), while for $M=12$ GeV it is around 100 (brown line).

\section{Discussion}\label{conclusions}

Scalar-tensor theories where the gravitational interaction is mediated by both the metric and scalar fields 
 arise commonly  in  modifications of standard general relativity theories.
The prototype example is the Brans-Dicke theory where the metric and the scalar field are related  via the conformal coupling as $\tilde g_{\mu\nu} =C(\phi) g_{\mu\nu}$.  However, the most general physically consistent relation between two metrics which can be given
by a single scalar field includes a disformal (or derivative) coupling \cite{Bekenstein}: $\tilde g_{\mu\nu} =C(\phi) g_{\mu\nu} + D(\phi) \partial_\mu\phi\partial_\nu\phi$.  

Both  couplings $C$ and $D$ can give rise to a different expansion rate from the standard cosmological model in the early universe, and still be in agreement with current constraints from BBN and gravity. In particular, BBN imposes a strong constraint on these couplings and it is encoded in the speed-up parameter $\xi$  (Eq.~\eqref{xi}), which needs to be very close to one before the onset of BBN.
It was shown in Ref.~\cite{Catena} that the expansion rate modification due to a conformal coupling can change the predictions on the dark matter relic abundances, anticipating freeze-out and producing a reannihilation phase for certain choices of initial conditions, as discussed in Ref.~\cite{DJZ}.

As was shown in Ref.~\cite{KWZ}, the (conformal and) disformal transformation naturally arises from  D-branes and thus in D-brane models of cosmology. In this case, the functions $C$ and $D$ are closely related and are dictated by the UV theory, for example, type IIB string theory. Moreover, the scalar field has a geometrical origin in terms of the transverse fluctuations of the D-brane, while  matter lives on the brane and it comes from the longitudinal fluctuations. 

In this paper we have studied the modification to the expansion rate due to the  disformally coupled scalar arising in D-brane-like models for cosmology, where $D=1/M^4 C$ (Figs.~\ref{HubbleC1},  \ref{HubbleScaleM} and \ref{HubbleCnon1}).  
Using the modified expansion rates thus found, we solved the  Boltzmann equation  to compute the  dark matter relic abundances (Fig.~\ref{Y100}). To find solutions, we used the current cosmological data on the DM content to determine the required thermally averaged cross sections (Fig.~\ref{Xsections}).  

We   numerically solved the coupled equations for $H$ and $\varphi$ (Eqs.~\eqref{Hprime1} and \eqref{phiHeq1}) and used this to  find the modified expansion rate. Note that contrary to the purely conformal case, in the presence of the disformal term it is not possible to eliminate $H$ from the system to solve a single master equation, as in Ref.~\cite{Catena}. So we need to carefully take into account both equations as well as the initial conditions for $H$. This introduces a cubic equation for $H$ in terms of the other parameters $(\varphi, \tilde\rho)$ and  a lower bound for the scale $M$, given the initial conditions for $(\varphi_i, \tilde\rho_i)$ (see section \ref{ICLowM}).

In section \ref{Ceq1} we presented for the first time the purely disformal case corresponding to $C=$ const. where the modification to the expansion rate is fully driven by the derivative coupling through $\gamma$ (see Eq.~\eqref{xiConst}). For this case, the modified expansion rate is always enhanced with respect to the standard one (Fig.~\ref{HubbleC1}), which implies an anticipated freeze-out and an enhancement of the cross section $\langle \sigma v\rangle$ (Fig.~\ref{Xsections}).
These results are robust against different initial conditions and we further studied the dependence on the scale $M$. We found that the larger the value of $M$, the earlier in the cosmological evolution the enhancement in the expansion rate (Fig.~\ref{HubbleScaleM}). Therefore, depending on the value of $M$, the modified expansion rate can appear at different times in the early universe  and the expansion rate can be $\sim 500$ times bigger compared to the standard GR case.  This will affect  any physical process in the early universe  where the Hubble expansion rate is needed to determine the out-of-equilibrium  temperature. In this paper we focused on the effect on the relic abundance and the annihilation rate of dark matter.  

Our results are also robust compared to the phenomenological disformal models usually discussed in the literature and  in Ref.~\cite{DJZ}. In that case, the conformal and disformal functions are unrelated (up to causality constraints). In that case, a purely disformal contribution will be obtained by setting $C=1$ and letting $D$  be an arbitrary function, fixed only by phenomenological constraints. The disformal enhancement in the expansion rate is similarly encoded in $\gamma$ in Eq.~\eqref{xiConst} and different choices for the function $D$ would be equivalent to different values of $M$ in the present paper. Therefore, our analysis is completely general and also applies to these phenomenological  models.

For the  $C\ne$ const case (section \ref{Cne1}) we saw that it is possible to have an enhancement as well as a reduction of the expansion rate with respect to the standard case, that is $\xi>1$ and $\xi<1$ (Fig.~\ref{HubbleCnon1}). This diminution gives the possibility of a reannihilation process, as in the conformal and disformal cases studied in Refs.~\cite{Catena,DJZ}. Thus the effect on the relic abundance and annihilation rate is analogous to the case studied in Ref.~\cite{DJZ}.   Again, we studied the effect of $M$ on the profiles of the expansion rates.  We considered in detail for concreteness only the  conformal function used in Refs.~\cite{Catena,DJZ}. For this function, the numerical analysis is relatively simple; however, there is in principle no obstruction to find similar effects for other functions. We looked at the function  $C= (b x^2 +c)^2$ (for  $b=4,8,15$, $c=1$), which can be a toy model for a smooth warp factor in a string theory setup. For this example, we found a relatively small enhancement  with $\xi\sim 4$.
 We expect that a  wider search of parameters and conformal functions will give rise to a larger enhancement as well as a decrease in the modified expansion rate.

\subsection*{Post-inflationary string cosmology}

The period after the end of inflation,  from reheating up to the onset of BBN remains largely  
unconstrained. 
 Let us now  connect our results with a post-inflationary toy model of string cosmology and discuss the implications in terms of the parameters of the theory.  

As described in section \ref{Sec:2}, we imagine a toy model where matter is coupled conformally and disformally to a scalar field, associated to an overall position of a (stack of) D-brane(s) in the internal six-dimensional compact space in a warped type IIB string compactification. In this case, we can relate the scale $M$ to the tension of a D3-brane  $T_3$ (for example), and thus to the string coupling $g_s$ and the string scale $M_s$ (or the six-dimensional volume ${\cal V}_6$) as $M=T_3^{1/4}= M_s (2\pi g_s^{-1})^{1/4}$.
The pure disformal case $C=1$ is very interesting and would correspond to a large-volume  compactification, where the warping due to the presence of fluxes can be ignored \cite{LV1,LV2}. 
In this case, the lowest value we used for $M$ that is relevant for the DM relic abundance was $\sim 10$ GeV and the largest (with a large effect) was $M\sim 300$. For string couplings of order $g_s\sim 10^{-4}$, these give string scales $\sim 1-20$ GeV and thus exponentially  large volumes ${\cal V}_6 \sim 10^{25}-10^{27}$,   that is,  a very low string scale and very weakly coupled compactification. 
%
On the other hand, for larger values of $M$, or larger values of the string scale, the enhancement on the expansion rate will occur earlier in the universe's evolution (see Fig.~\ref{HubbleScaleM}). For example,  the largest value we used, $M\sim 10^{10}$  GeV, for $g_s\sim 10^{-4}$ would give $M_s\sim 10^{9}$ GeV, volumes of order ${\cal V}_6\sim 10^{10}$ and the expansion enhancement occurs at around $\tilde T\sim 10^{10}$ GeV. 
Therefore, depending on the string scale, coupling and compactification, we may expect the early pre-BBN cosmology to be affected at different epochs with interesting consequences for the post-inflationary string cosmological evolution. This will also be connected to string inflation, which usually requires large string scales (see Ref.~\cite{BM} for a review).

Of course a more  realistic model may involve,  for example, other parameters of the theory in the scale $M$ (due for example to higher-dimensional D-branes wrapping the internal space), nonuniversal couplings among the scalar (or scalars) to SM matter and DM as briefly discussed in Ref.~\cite{MW} for the pure conformal case, etc. However, we find it very  interesting that scalar couplings present in string theory can give important predictions for the post-inflationary evolution after string inflation. 

Let us finally stress that the cosmological implications of conformal and disformal couplings  in scalar-tensor theories are in any case very interesting  from a more phenomenological point of view. Here we have taken a further step in making progress to address these implications and have presented the disformal effects 
 for the first time.

\acknowledgments 
We would like to thank Enrico Nardi and Gianmassimo Tasinato for useful discussions and Gianmassimo Tasinato for comments on the manuscript.  BD and EJ acknowledge support from DOE Grant de-sc0010813. IZ is partially supported by the STFC grants ST/N001419/1 and ST/L000369/1. IZ would like to thank the Mitchell Institute for Fundamental Physics and Astronomy at Texas A\&M University for kind hospitality while part of this work was done. Part of this work  was performed at the Aspen Center for Physics, which is supported by National Science Foundation grant PHY-1607761.

\bibliographystyle{utphys}
\bibliography{refsDBrane}

\end{document}